\documentclass[twocolumn,tighten]{aastex63}

\usepackage{graphics,graphicx}
\usepackage{epsfig}
\usepackage{color}
\usepackage{amsmath}
\usepackage{ulem}   
\newcommand{\ltsimeq}{\raisebox{-0.6ex}{$\,\stackrel
        {\raisebox{-.2ex}{$\textstyle <$}}{\sim}\,$}}
\newcommand{\gtsimeq}{\raisebox{-0.6ex}{$\,\stackrel
        {\raisebox{-.2ex}{$\textstyle >$}}{\sim}\,$}}

\newcommand{\chemone}{\raisebox{0.03cm}{$-$}} 
\newcommand{\chemtwo}{\raisebox{0.03cm}{$=$}} 

\usepackage{xcolor,soul}

\received{2020.Aug.23}
\revised{2020.Nov.09}
\accepted{2020.Nov11}
\submitjournal{The Planetary Science Journal }

\shortauthors{Woodward et al.}

\begin{document}

\title{The Coma Dust of Comet C/2013 US$_{10}$ (Catalina) -- A Window into Carbon in the Solar System}

\correspondingauthor{C.E. Woodward}
\email{chickw024@gmail.com}

\author{Charles E. Woodward}
\affiliation{Minnesota Institute for Astrophysics, University of Minnesota,
116 Church Street SE, Minneapolis, MN 55455, USA}
\affiliation{Visiting Astronomer at the Infrared Telescope Facility, which is
operated by the University of Hawaii under contract\\ 80HGTR19D0030
with the National Aeronautics and Space Administration.}

\author{Diane H. Wooden}
\affiliation{NASA Ames Research Center, MS 245-3, Moffett Field,
CA 94035-0001, USA}

\author{David E. Harker}
\affiliation{University of California, San Diego, Center for
Astrophysics \& Space Sciences,\\ 9500 Gilman Dr. Dept. 0424, La Jolla,
CA 92093-0424, USA}

\author{Michael S. P. Kelley}
\affiliation{University of Maryland, Department of Astronomy,
College Park, MD 20742-2421, USA}
\affiliation{Visiting Astronomer at the Infrared Telescope Facility, which is
operated by the University of Hawaii under contract\\ 80HGTR19D0030
with the National Aeronautics and Space Administration.}

\author{Ray W. Russell}
\affiliation{The Aerospace Corporation, P.O. Box 92957, M2-266, 
Los Angles, CA 90009, USA}
\affiliation{Visiting Astronomer at the Infrared Telescope Facility, which is
operated by the University of Hawaii under contract\\ 80HGTR19D0030
with the National Aeronautics and Space Administration.}

\author{Daryl L. Kim}
\affiliation{The Aerospace Corporation, P.O. Box 92957, M2-266,
Los Angles, CA 90009, USA}
\affiliation{Visiting Astronomer at the Infrared Telescope Facility, which is
operated by the University of Hawaii under contract\\ 80HGTR19D0030
with the National Aeronautics and Space Administration.}

\begin{abstract}

Comet C/2013 US$_{10}$ (Catalina) was an dynamically new Oort cloud comet whose
apparition presented a favorable geometry for observations near close Earth
approach ($\simeq 0.93$ au)  at heliocentric distances $\ltsimeq 2$~au when
insolation and sublimation of volatiles drive maximum activity. Here we present
mid-infrared $6.0 \ltsimeq \lambda(\mu\rm{m}) \ltsimeq 40$ spectrophotometric
observations at two temporal epochs from NASA's Stratospheric Observatory 
for Infrared Astronomy and the NASA Infrared Telescope Facility 
that yield an inventory of the refectory materials 
and their physical characteristics through thermal modeling analysis.  
The grain composition is dominated by dark dust grains (modeled as amorphous 
carbon) with a silicate-to-carbon ratio $\ltsimeq 0.9$, little of crystalline 
stoichiometry (no distinct 11.2~\micron{} feature attributed to Mg-rich 
crystalline olivine), the submicron grain size distribution peaking at 
$\simeq 0.6$ \micron.  The 10~\micron{} silicate feature was weak,
$\approx 12.8 \pm 0.1$\% above the local continuum, and the bolometric grain 
albedo was low ($\ltsimeq 14$\%). Comet C/2013 US$_{10}$ (Catalina) is 
a carbon-rich object. This material, which is well-represented by the 
optical constants of amorphous carbon is similar to the material that darkens and 
reddens the surface of comet 67P/Churyumov-Gerasimenko. We argue 
this material is endemic the nuclei of comets, synthesizing results from the study 
of \textit{Stardust} samples, interplanetary dust particle investigations and 
micrometeoritic analyses. The atomic carbon-to-silicate ratio of 
comet C/2013 US$_{10}$ (Catalina) and other comets joins a growing body of evidence
suggesting the existence of a C/Si gradient in the primitive solar system, providing new
insight to planetesimal formation and the distribution of isotopic and compositional 
gradients extant today.

\end{abstract}




\section{Introduction} \label{sec:intro}

Traces of primordial materials, and their least-processed products,
are to be found in the outermost regions of the solar system in the form of 
ices of volatile materials (H$_{2}$O, CO, CO$_{2}$, and other
more rare species), and more refractory dust grains. This is the realm
of comets. Nevertheless, it is certain that
this outer region beyond the frost line was not entirely
``primordial'' but was ``polluted'' with the processed
materials from the inner disk, the ``hot nebular products,''
\citep{2014AREPS..42..179B, 2011ApJ...740....9C, 2007prpl.conf..815W, 2002ApJ...565L.109H},
where gas-gas and gas-grain reactions occurred
\citep{2017A&A...606A..16G, 2004A&A...413..571G, 2002A&A...390..253G}. There
is considerable evidence that in the cold regions where cometary 
material formed, forming comet bodies were ``salted'' with refractory 
material processed at much higher temperatures \citep{2006Sci...314.1735Z}.

Considerable efforts have been expended to
characterize the nature of refractory cometary grains to
understand the environment of the early solar system from pebbles to
planetesimals to larger bodies \cite[see][and references therein]{2016MNRAS.462S..23P}. 
These grains likely are minimally processed over the age of the solar system after 
incorporation into the nuclei of comets. Information on the nature of these 
grains comes from a variety of sources, including remote sensing 
through telescopic observations (ground-based, airborne, and
space-based), rendezvous/encounter experiments
(i.e., \textit{Giotto, Rosetta/Philae, Deep Impact}), collection
of interplanetary dust particles (IDPs) in the Earth's stratosphere,
and a sample return mission (\textit{Stardust}). All these activities have made
important contributions to our understanding of these grains. The
most detailed information we have comes from the latter two types
of studies, where laboratory analysis is possible. Yet, the
IDPs from comets 81P/Wild 2 and 26P/Grigg-Skjellerup are
vastly different. The former contains material processed at
high temperature \citep{2006Sci...314.1735Z} while the latter is
very ``primitive'' \citep{2009E&PSL.288...44B}. For these reasons, it
is necessary to determine as best we can the properties of dust
grains from a large sample of comets using remote
techniques \citep{2015SSRv..197....9C}. These include observations of 
both the thermal (spectrophotometric) and scattered light
(spectrophotometric and polarimetric). The former technique provides
our most direct link to the composition (mineral content) of the grains.

With these data, combined with modeling features in the infrared spectral
energy distribution (SEDs) arising from mineral species
emitting in the comet coma (dust grains) and dynamical models of solar system 
formation and planetary migration we can address fundamental questions of
solar system formation. These question include: What was the method 
of transport of these materials, and has information on the scale of those 
transport processes been stored in primitive solar system objects? Do comets, 
the remnants of that epoch, still contain clues as to what happened?

In this paper we report our post-perihelion (TP = 2015 Nov 15.721~UT)
spectrophotometric observations of comet C/2013 US$_{10}$ 
(Catalina), a dynamically new
\citep[see][for a definition based on orbital elements]{1950BAN....11...91O}
Oort Cloud comet with  $1/a_{org} = 5.3 \times 10^{-6}$~AU$^{-1}$
\citep{mpec2019-J142} and discuss important new interpretations that the
coma grain composition of comets from remote sensing observations can
bring to understanding disk processing in the primitive solar system.

\newpage
\section{Observations}\label{sec:obsec}

Infrared and optical observations of C/2013 US$_{10}$ (Catalina) were conducted at two 
contemporaneous epochs near close Earth approach ($\Delta \simeq 0.93$~au) with 
the NASA Infrared Telescope Facility (IRTF) and NASA's Stratospheric Observatory 
for Infrared Astronomy (SOFIA) facility. Table~\ref{tab:sobstab_tab} summarizes the all 
observational data sets discussed herein and physical parameters of the comet.


\subsection{Ground-based Spectrophotometry}\label{sec:obs-bass}

Medium resolution ($R \equiv \lambda/\Delta\lambda \simeq 50-120$)
infrared spectroscopy of comet C/2013 US$_{10}$ (Catalina) was obtained on 
the NASA IRTF telescope with The Aerospace Corporation's Broadband Array
Spectrograph System \citep[BASS;][]{1990SPIE.1235..171H} during the early morning 
(daytime) hours. BASS has no moving parts and observes all wavelengths in 
its 2 to 14~\micron{} operable range using two 58 element block impurity band 
linear arrays simultaneously through the same aperture. All observations 
were obtained with a fixed 4\farcs0  diameter circular aperture.
Standard infrared observing techniques were employed, using double beam mode 
with a chop/nod throw of $\simeq 60$\arcsec. \citet{2002M&PS...37.1255S} provide a 
detailed description of the BASS data acquisition and preliminary reduction scheme. 
Non-sidereal tracking of the comet by the IRTF telescope was performed using 
Jet Propulsion Laboratory (JPL) Horizons' \citep{1996DPS....28.2504G} generated rates, 
and fine guiding, to keep the comet photocenter
in the BASS aperture, was done either by manually guiding on the visible comet image
produced by the BASS sky-filtered visible CCD camera, or off a strip-chart
using thermal channels of the BASS array.

Photometric calibration of individual comet data sets were performed using 
observations of $\alpha$~Boo observed at equivalent airmass to minimize telluric 
corrections. $\alpha$~Boo is a well-characterized infrared standard for ground- and space-based 
telescopes and has been extensively monitored and modeled by the BASS instrument team 
and other investigators for decades. The calibration and telluric corrections 
are uncertain to within $\simeq3$\%. Examination of independent, flux 
calibrated spectra of comet C/2013 US$_{10}$ (Catalina) obtained during the course of 
the 2016 Jan 10.61~UT observational campaign showed no variance in the flux 
level of the spectral energy distribution (i.e., no outbursts, or jet induced 
changes in coma brightness were witnessed), or spectral shape. Hence, all 
spectra where averaged 

\begin{figure}[!ht]
\figurenum{1}
\includegraphics[trim=1.6cm 0cm 0cm 1.60cm,clip,width=1.10\columnwidth]{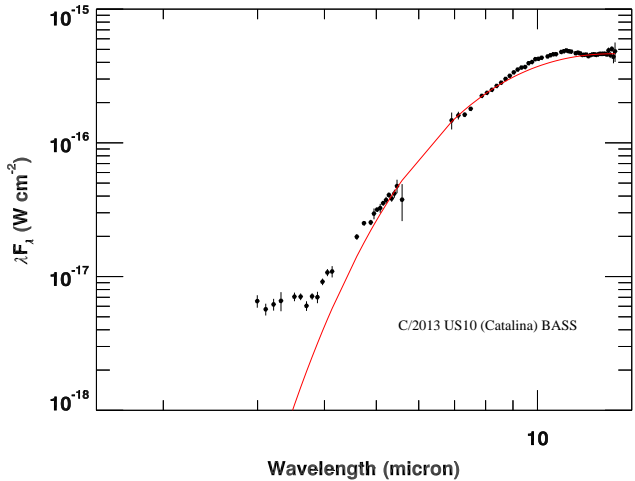}
\caption{Comet C/2013 US$_{10}$ (Catalina) 3.0 to 14~\micron{} BASS spectrum obtained on 
2016 Jan 10.61 UT with the NASA IRTF telescope. This spectrum was derived by 
averaging all photometrically calibrated individual comet spectra obtained over a 1.33~hr 
interval. Regions of poor telluric transmission ($\ltsimeq 30$\%) where from atmospheric CO$_{2}$ and 
H$_{2}$O vapor have strong absorption bands result in gaps in the data where BASS spectral 
data points are clipped out. The red curve is the best-fit blackbody, T$_{\rm{BB}}
= 265.3\pm 2.6$~K fit to the local 10~\micron{} continuum as described in \S\ref{sec:mir_analysis}.
The excess over the blackbody curve at short wavelengths is due to scattered, reddened sunlight 
contributing substantially to the flux.
\label{fig:obsflx_bass_20160110}}
\end{figure}

\begin{deluxetable*}{@{\extracolsep{0pt}}lccccccccc}
\tablenum{1}
\setlength{\tabcolsep}{2pt} 
%
%
\tablecaption{Observational Summary -- Comet C/2013 US10 (Catalina)\tablenotemark{a}\label{tab:sobstab_tab}}
\tablehead{
\colhead{Mean}&&\colhead{Grism} &\colhead{Single}&\colhead{Total} &&&&\colhead{Tail\tablenotemark{b}} &\colhead{Tail\tablenotemark{b}}\\
\colhead{Observation}&&\colhead{or}&\colhead{Frame}&\colhead{On Source}&&&&\colhead{anti-Sun} &\colhead{anti-velocity}\\
\colhead{Date} & \colhead{Instrument} & \colhead{Filter} & \colhead{Exposure} &\colhead{Integration} &&& \colhead{Phase} & \colhead{vector} & \colhead{vector}\\
\colhead{2016 UT} &\colhead{Configuraion} &\colhead{$\lambda_{c}$} &\colhead{Time} &\colhead{Time} &\colhead{$r_{h}$} &\colhead{$\Delta$} &\colhead{Ang} &\colhead{Gas} &\colhead{Dust}\\
\colhead{(dd-mm hr:min:s)} &  &\colhead{($\mu$m)} &\colhead{(sec)}  &\colhead{(sec)} &\colhead{(AU)} &\colhead{(AU)} &\colhead{($\circ$)} &\colhead{($\circ$)} &\colhead{($\circ$)}
 }

\startdata
\underbar{NASA SOFIA}\\[2pt]
FORCAST (FOF276)\\
02-10T07:06:59.6 & Imaging SWC  & \phn7.70  & 23.88 & 477.52 & 1.710 & 1.106 & 32.96 & 103.54 & 23.31\\
02-10T07:29:47.6 & Imaging Dual & 11.01 & 18.79 & 244.28 & \nodata & \nodata & \nodata & \nodata & \nodata\\
02-10T07:46:06.4 & Imaging Dual & 19.70 & 21.38 & 171.05 & \nodata & \nodata & \nodata & \nodata & \nodata\\
02-10T07:29:47.6 & Imaging Dual & 31.36 & 19.29 & 405.17 & \nodata & \nodata & \nodata & \nodata & \nodata\\
02-10T08:01:00.8 & Grism SWC    & G063  & \phn5.00  & 669.50 & \nodata & \nodata & \nodata & \nodata & \nodata\\[2pt]
FORCAST (FOF275)\\
02-09T08:09:02.3 & Imaging SWC  & \phn7.70  & 24.90 & 498.07 & 1.697 & 1.080 & 33.06 & 106.50 & 24.80\\
02-09T08:31:48.8 & Imaging Dual & 11.01 & 18.83 & 131.08 & \nodata & \nodata & \nodata & \nodata & \nodata\\
02-09T08:38:42.9 & Imaging Dual & 11.01 & 18.83 & 131.08 & \nodata & \nodata & \nodata & \nodata & \nodata\\
02-09T08:46:50.0 & Imaging Dual & 19.70 & 20.05 & 180.49 & \nodata & \nodata & \nodata & \nodata & \nodata\\
02-09T08:31:48.8 & Imaging Dual & 31.36 & 18.83 & 131.80 & \nodata & \nodata & \nodata & \nodata & \nodata\\
02-09T08:38:42.9 & Imaging Dual & 31.36 & 18.83 & 131.80 & \nodata & \nodata & \nodata & \nodata & \nodata\\
02-09T09:07:56.2 & Grism LWC    & G111  & 11.82 & 574.73 & \nodata & \nodata & \nodata & \nodata & \nodata\\
02-09T09:52:30.2 & Grism LWC    & G227  & 12.00 & 816.07 & \nodata & \nodata & \nodata & \nodata & \nodata\\
02-09T10:34:58.3 & Grism LWC    & G329  & 11.97 & 742.35 & \nodata & \nodata & \nodata & \nodata & \nodata\\[3pt]
\underbar{NASA IRTF}\\[2pt]
MORIS\\
01-11T15:11:06.6 & Imaging Optical  & SDSS $i^{\prime}$  & 5.00 & 20.00 & 1.315 & 0.747 & 47.80 & 292.10 & 157.27\\[2pt]
BASS\\
01-10T14:34:30.0 & IR Spectra & 2.6-14.2 & 960.00 & 4800.00 & 1.307 & 0.758 & 48.72 & 293.42 & 157.37\\
\enddata
\tablecomments{}
\tablenotetext{a}{Observation geometry calculated by JPL Horizons \citep{1996DPS....28.2504G}.}
\tablenotetext{b}{Vector direction measured CCW (eastward) from celestial north on the plane of the sky.}

\end{deluxetable*}



\noindent together (with the proper propagation of all 
statistical point-to-point uncertainties) to produce the final spectrum
presented in Fig.~\ref{fig:obsflx_bass_20160110}.

Optical imagery of the comet was obtained on 2016 January 11.633~UT
with the NASA IRTF MORIS camera \citep{2011PASP..123..461G} in
a Sloan Digital Sky Survey (SDSS) $i^{\prime}$ filter 
($\lambda_{c} = 0.7630$~\micron{}, filter full width half maximum (FWHM) of 
0.1530~\micron). Multiple exposures (5 sec each) of the comet nucleus and 
surrounding coma were obtained using AB pairs nodding the telescope by 60\arcsec, 
and dithering the telescope while tracking at the non-sidereal rate corresponding 
to the predicted motion of the comet in an airmass range of $\approx 1.18$. All 
images were corrected for overscan, and bias with standard IRAF\footnote{IRAF is 
distributed by the National Optical Astronomy Observatory, which is
operated by the Association of Universities for Research in Astronomy (AURA) 
under cooperative agreement with the National Science Foundation.} routines. 
The data was photometrically calibrated using GSC 02581--02323 (G2V) SDSS colors 
reported from SIMBAD transformed to the USNO system as described 
in \citet{2006AN....327..821T}, adopting 3631 Jy for zeroth magnitude.
No color corrections for spectral type were applied in the transformation. The average nightly 
seeing was $\sim$~2\farcs2 as determined from the standard star. The observed $i^{\prime}$ flux 
density of the comet measured in an equivalent BASS aperture was 
$(2.316 \pm 0.001) \times 10^{-17}$~W~cm$^{-2}$~\micron$^{-1}$. 

\begin{figure}[!ht]
\figurenum{2}
\includegraphics[trim=1.8cm 0.5cm 0cm 2.0cm,clip,width=1.10\columnwidth]{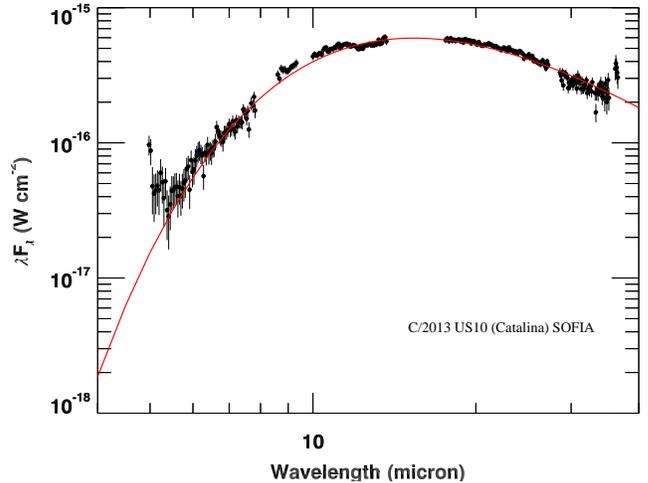}
\caption{\label{fig:obsflx_sofia_20160209} Comet C/2013 US$_{10}$ (Catalina) 
composite 4.9 to 36.5~\micron{} FORCAST spectrum 
obtained with SOFIA on 2016 Feb 09.41 UT (mean UT of both nights). Regions of poor 
telluric transmission ($\ltsimeq 70$\%) at flight altitudes result in gaps within certain wavelength
intervals within each individual grism where FORCAST spectral data points are clipped out. 
The red curve is the best-fit blackbody that
yields a  T$_{\rm{BB}} = 239.5\pm 0.5$~K fit using all wavelengths $\gtsimeq 6.0~\mu$m}
as described in \S\ref{sec:mir_analysis}. 
\end{figure}


\subsection{Airborne SOFIA Observations}\label{sec:obs-gensofia}

Mid-infrared (mid-IR) spectrophotometric observations of 
comet C/2013 US$_{10}$ (Catalina) were 
obtained using the Faint Object InfraRed CAmera \cite[FORCAST;][]{2018JAI.....740005H} mounted 
at the Nasmyth focus of the 2.5-m telescope of the SOFIA Observatory 
\citep{2012ApJ...749L..17Y}. FORCAST is a dual-channel mid-IR 
imager and grism spectrometer operating from 5 to 40~\micron. 

The data were acquired on two separate, back-to-back flights, originating from
Palmdale, CA at altitudes of $\simeq 11.89$~km in 2016 February, conducted as part 
of our SOFIA comet programs (P.I. Woodward, AOR\_ID 04\_0010). Mid-infrared imaging 
observations of C/2013 US$_{10}$~(Catalina) in three filters and the The Short Wavelength 
Camera (SWC) grism (G063) were obtained on the first flight, while on the second flight, 
imaging in the same three filters was repeated in addition to Long Wavelength Camera (LWC) 
grism observations with three gratings (G111, G227, and G329). For all spectroscopic observations 
the instrument was configured using a long-slit (4\farcs7 $\times$ 191\arcsec) 
which yields a spectral resolution $R = \lambda/\Delta\lambda \sim$ 140--300. The 
comet was imaged in the SWC using the F197 filter to position the target in 
the slit. Both imaging and spectroscopic data were obtained using a 2-point 
chop/nod in the Nod-Match-Chop (C2N) mode with 45\arcsec\, chop and 
90\arcsec\,nod amplitudes at angles of 30$^{\circ}$/210$^{\circ}$ in the 
equatorial reference frame. 

The FORCAST scientific data products were retrieved from the SOFIA archive, 
after standard pipeline processing and flux calibration was performed 
\citep[for details see][]{2015ASPC..495..355C, 2015ApJ...809..181W}.
An extensive discussion of the FORCAST data pipeline can be found in the Guest 
Investigator Handbook for FORCAST Data Products,
Rev. B\footnote{\url{https://www.sofia.usra.edu/Science/DataProducts/FORCAST\_GI\_Handbook\_RevA1.pdf}}

The computed atmospheric transmission at flight altitudes was used to clip-out 
grism data points in wavelength regions where the transmission was less than 
70\%. Subsequently, to increase the signal-to-noise (SNR) ratio of the comet 
spectra, data in each grism spectra segment were binned using a weighted 3-point 
boxcar. As there is no wavelength overlap between individual FORCAST grism 
segments, combined with an inherent uncertainty in the absolute 
grism flux calibration, and the fact that observations were conducted on separate 
nights, photometry derived from the image data was used to scale the grism data 
to a common spectral energy distribution (SED). Integration of the observed grism 
data with the corresponding filter transmission profile lying within the respective 
grism spectral grasp (i.e., FORF111 for G111) was used to construct a synthetic 
photometric point. This latter photometric point was compared to the observed image 
aperture photometry derived within an equivalent circular diameter beam corresponding 
to the grism extraction aperture area (average for all grisms was 
17\farcs54 $\pm$ 0\farcs74, derived data product keyword PSFRAD). The grism scaling factor 
was derived from this ratio ($\ltsimeq 8$\%). Neither the shape of the observed SED inferred from 
the image photometry nor the relative flux level of the SED changed significantly over 
the two epoch of the SOFIA observations. 

\begin{figure*}[!ht]
\figurenum{3}
\gridline{
\rotatefig{0}{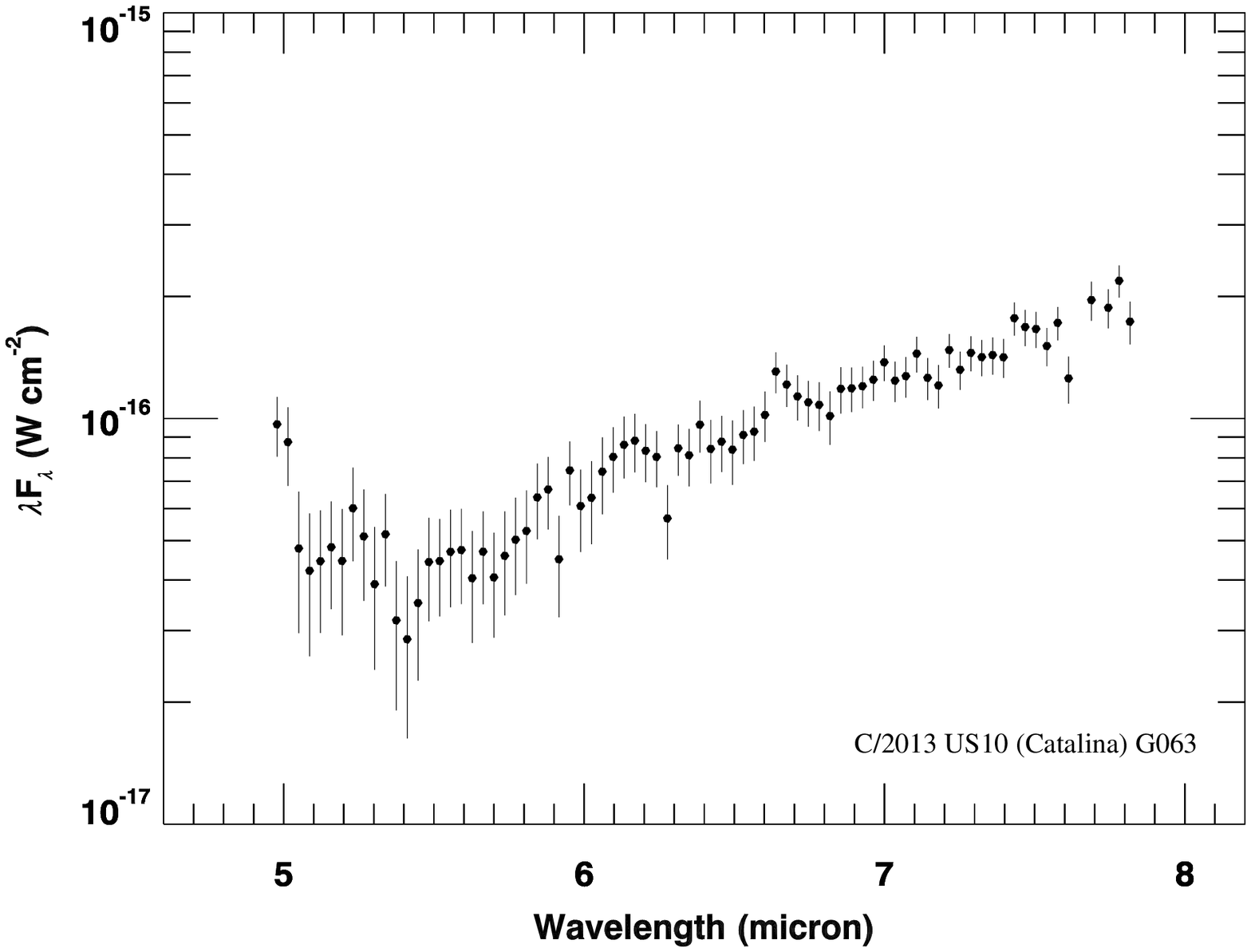}{0.9\columnwidth}{(a)}
\rotatefig{0}{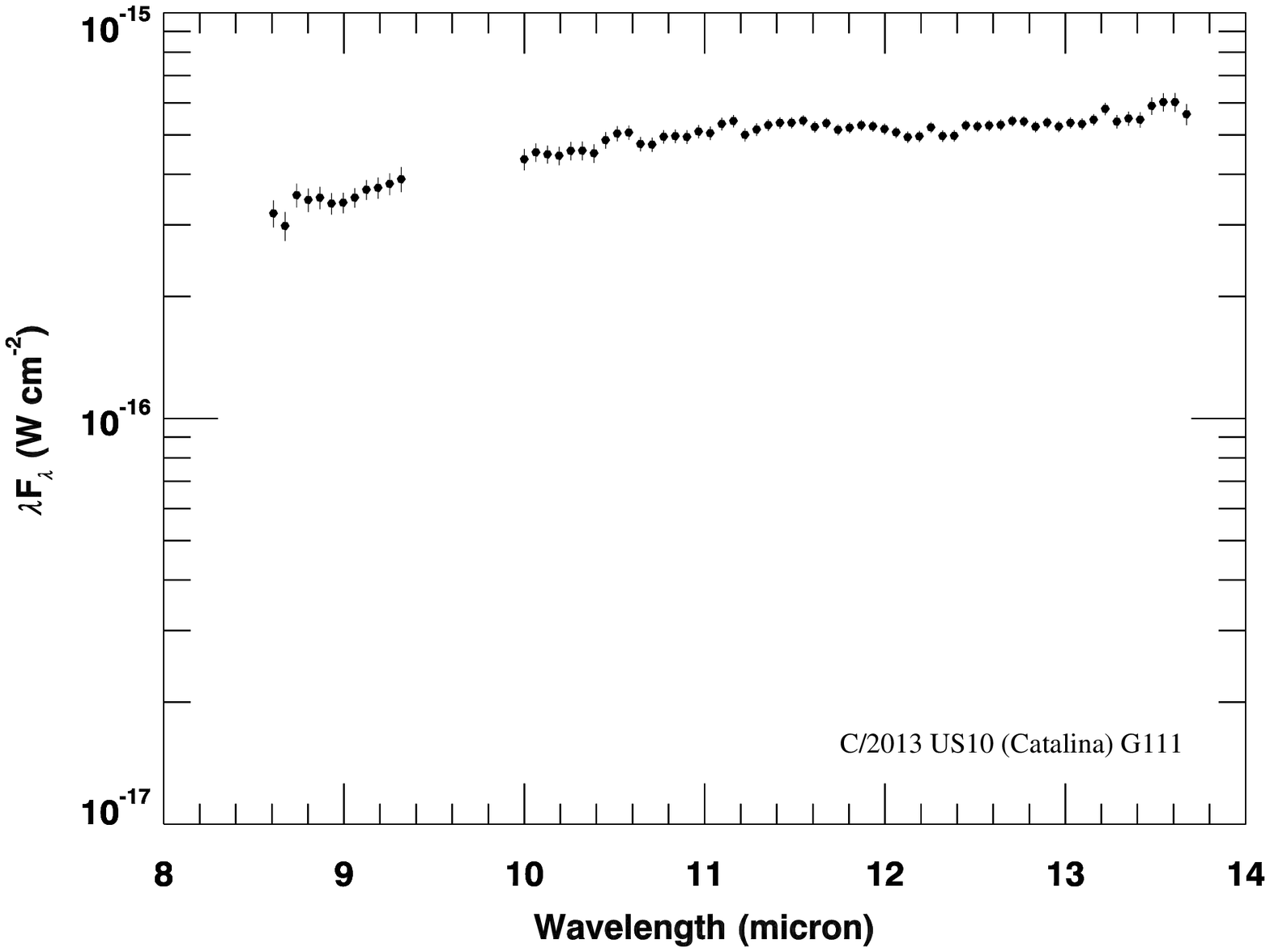}{0.9\columnwidth}{(b)}
}
\gridline{
\rotatefig{0}{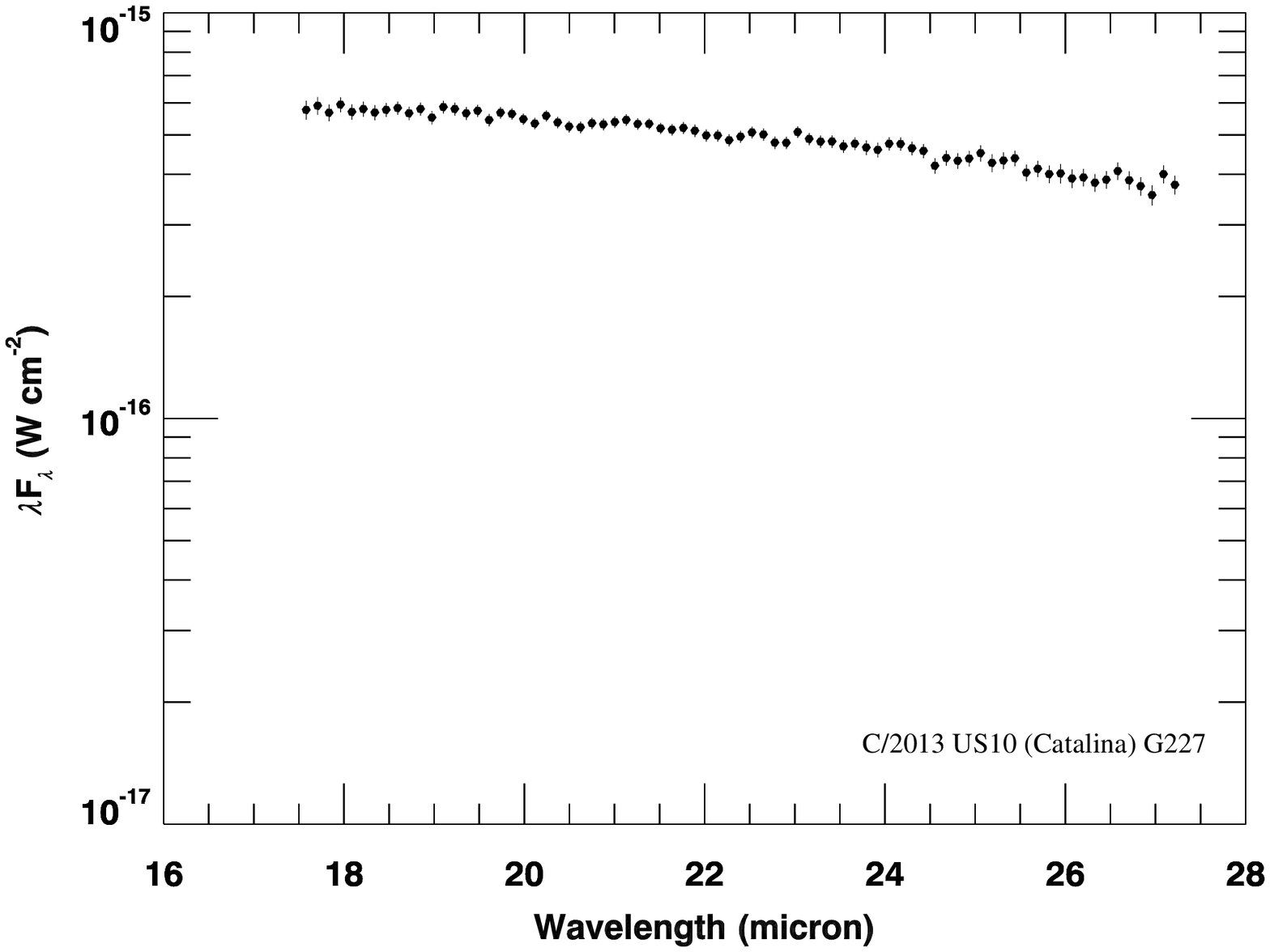}{0.9\columnwidth}{(c)}
\rotatefig{0}{fig3d}{0.9\columnwidth}{(d)}
         }
\caption{\label{fig:obs_each_grating_alone} Comet C/2013 US$_{10}$ 
(Catalina) SOFIA FORCAST spectra by individual grating to highlight spectral details 
and the signal-to-noise quality of the data. The panels are (a) G063, (b) G111, (c) G227, 
and (d) G329. The original spectra have been binned with a 3-point width 
(in wavelength-space) median boxcar, with the errors propagated by use of a weighted mean. 
Gaps in the contiguous spectral coverage arise from regions where the atmospheric transmission 
was modeled to be $\ltsimeq 70$\%.
}
\end{figure*}

The resultant composite FORCAST spectra of comet 
C/2013 US$_{10}$ (Catalina) is presented in Fig.~\ref{fig:obsflx_sofia_20160209}.
Figure~\ref{fig:obs_each_grating_alone} presents panels for each individual 
grating segment, spanning the respective spectral grasp, to illustrate spectral details
of the observed SEDs. 

Optical images in the SDSS $i^{\prime}$ filter also were obtained on each 
flight series prior to the start of the mid-infrared observing sequence using 
the Focal Plane Imager \citep[FPI+;][]{2016SPIE.9908E..2WP}. The FPI+ field-of-view 
is 8.7 square arcminutes, with a plate scale of 0\farcs51 per pixel, 
and a FWHM of $\simeq$3\farcs75.
The comet was tracked using the JPL Horizons non-sidereal rates. These data frames 
were bias and overscanned corrected using standard routines. The comet's surface 
brightness was flux calibrated by using aperture photometry of seven stars 
in the image field of view with known $i^{\prime}$ magnitudes taken from the 
USNO UCAC4 catalog to establish the photometric zero point 
(resultant fractional uncertainty of $\simeq 1$\%).  The observed $i^{\prime}$ flux density
of the comet measured in an equivalent circular aperture corresponding to the average SOFIA 
FORCAST grism extraction aperture was $(8.215\pm 0.009) \times 10^{-17}$~W~cm$^{-2}$~\micron$^{-1}$. 


\section{DISCUSSION}\label{sec:discuss}

\subsection{SOFIA Imagery and Photometry}\label{sec:obs-imagesofia}

\begin{figure*}[!ht]
\figurenum{4}
\gridline{
\rotatefig{90}{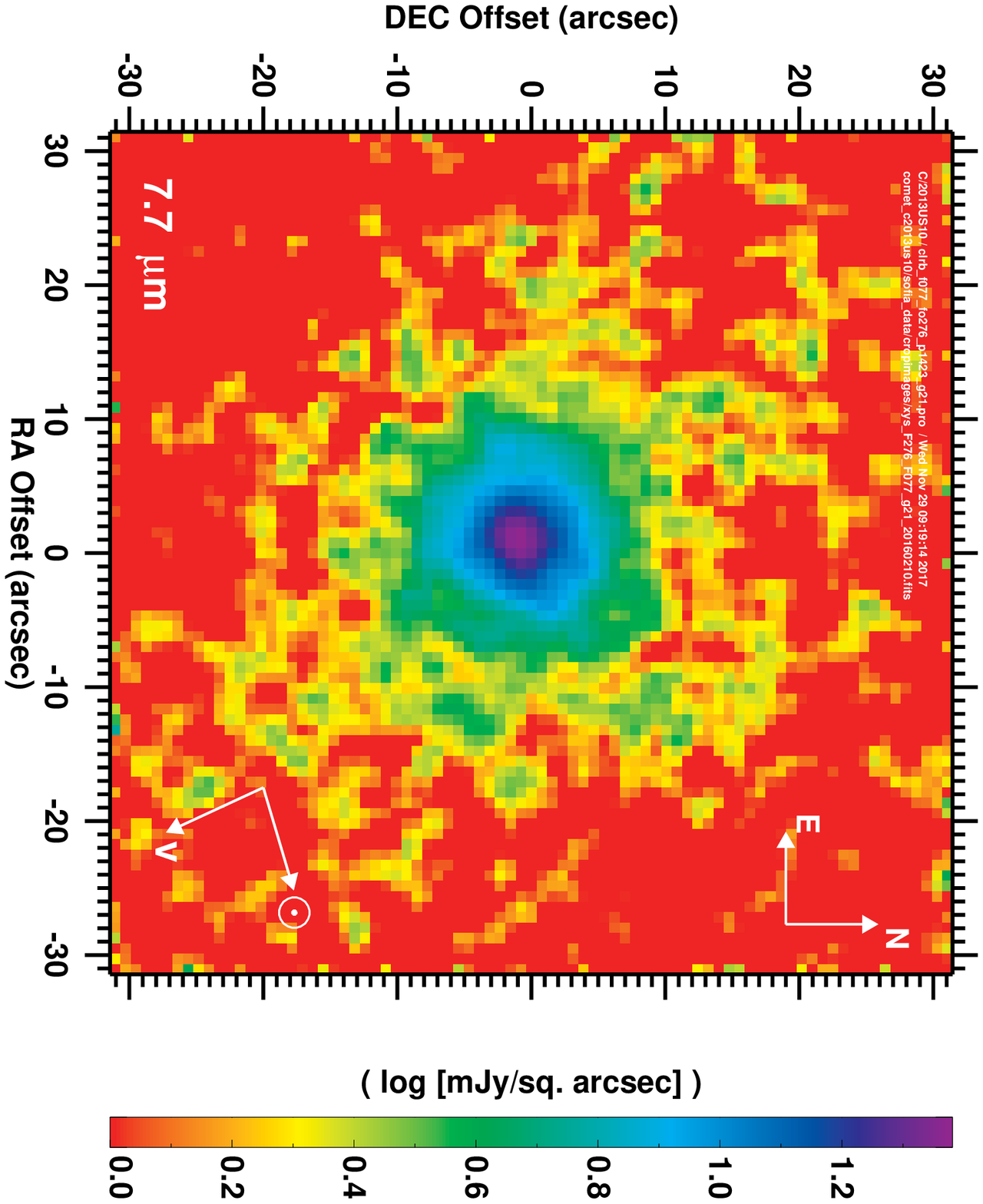}{0.9\columnwidth}{(a)}
\rotatefig{90}{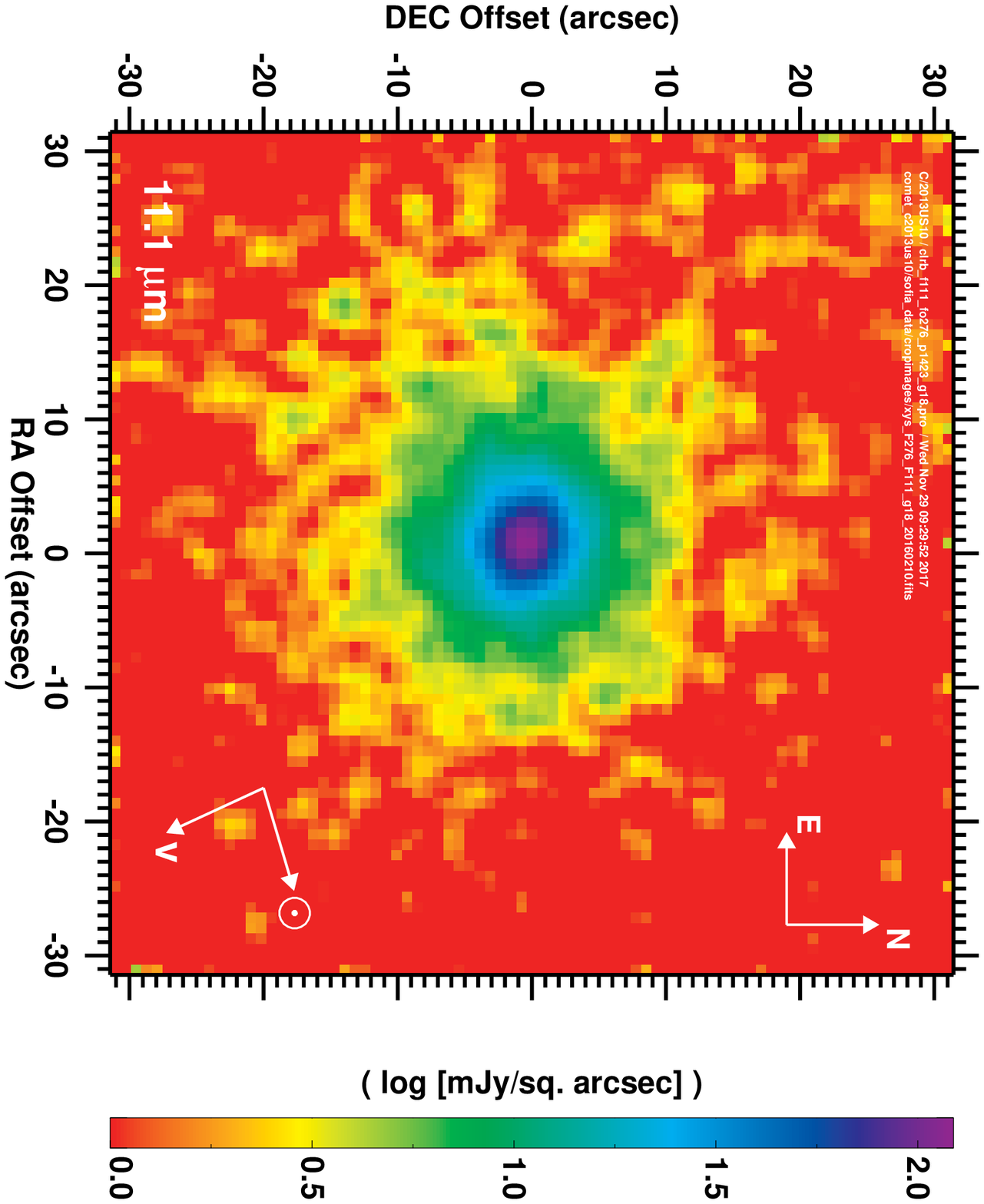}{0.9\columnwidth}{(b)}
         }
\gridline{
\rotatefig{90}{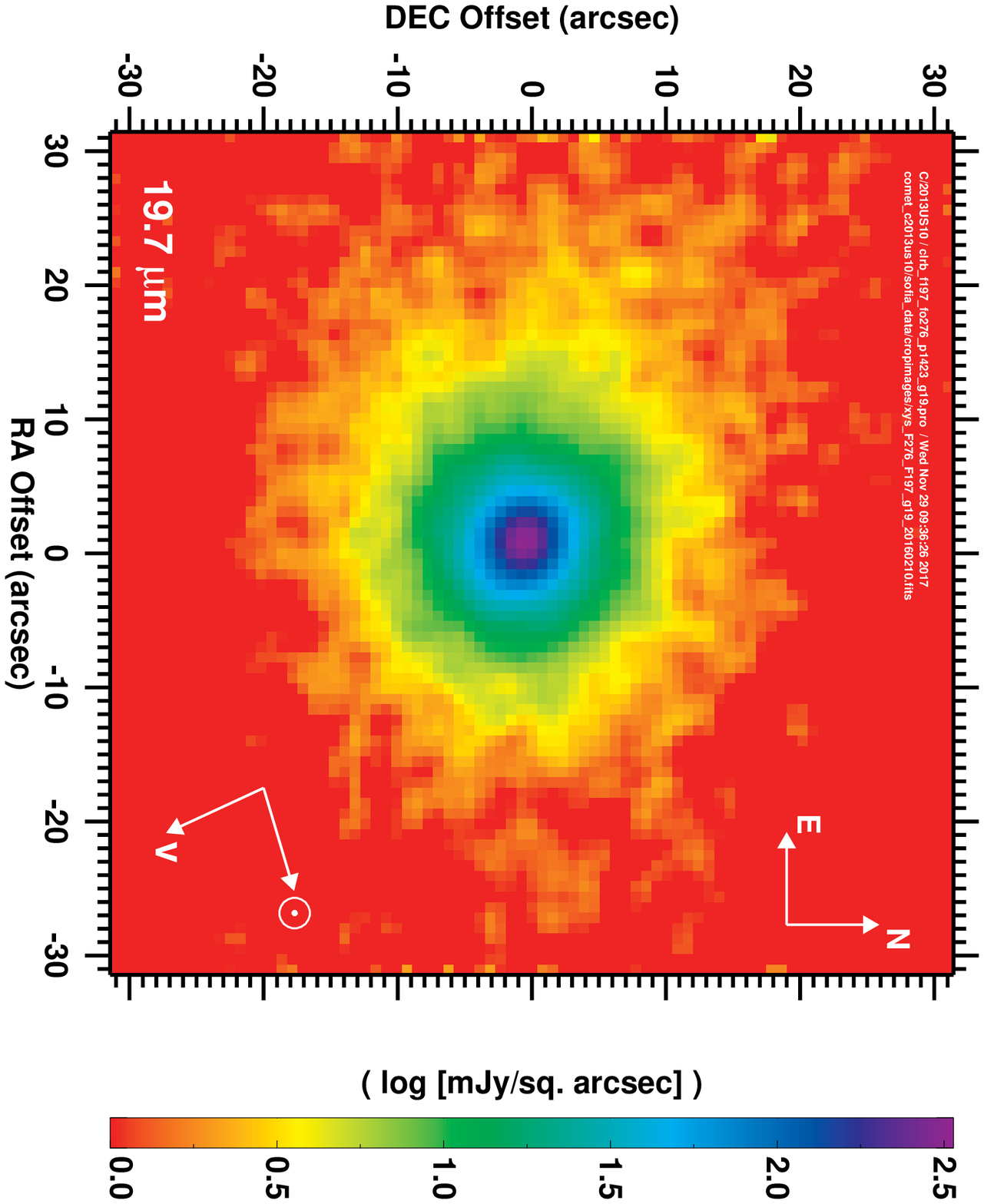}{0.9\columnwidth}{(c)}
\rotatefig{90}{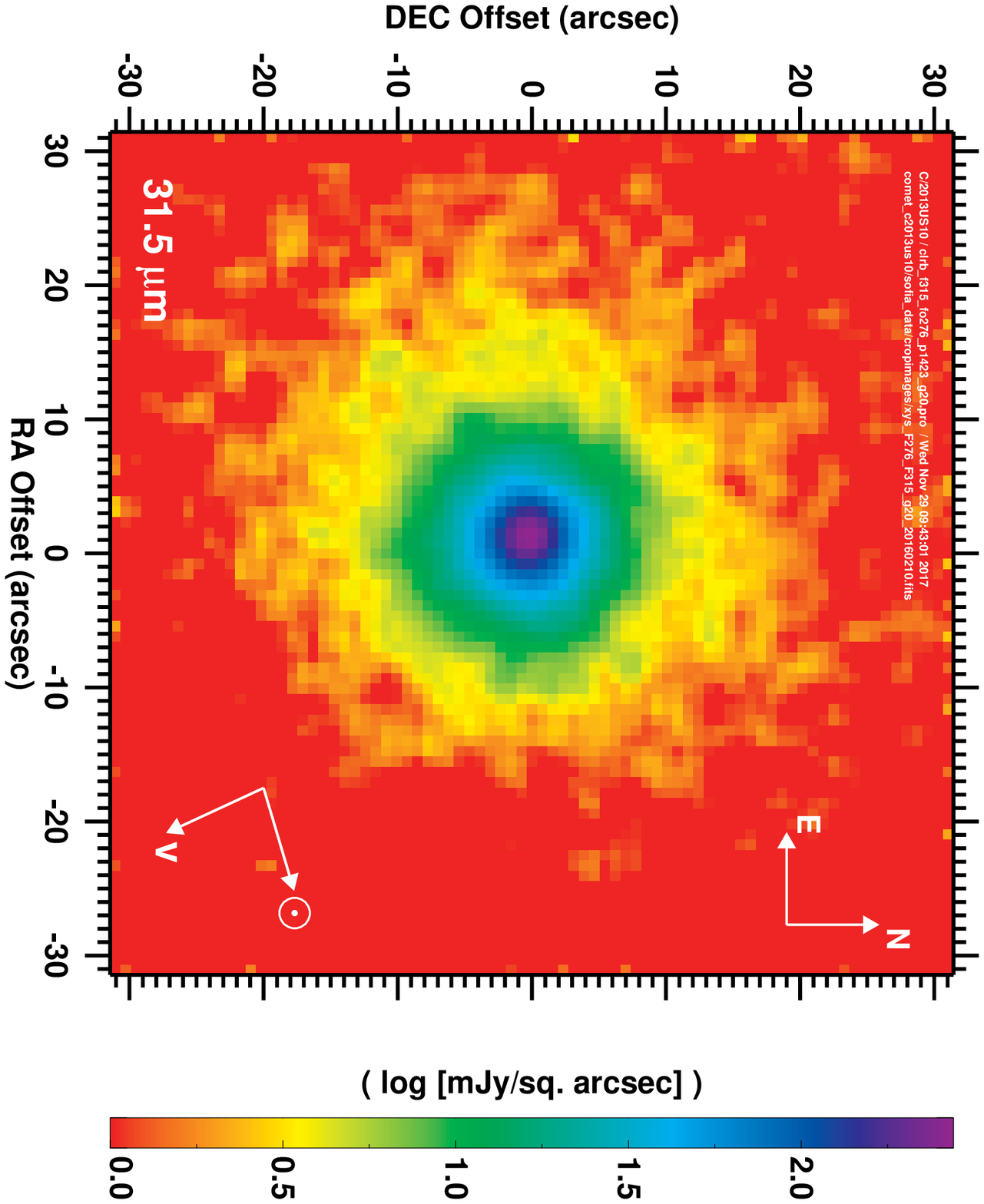}{0.9\columnwidth}{(d)}
         }
\caption{Comet C/2013 US$_{10}$ (Catalina) FORCAST filter imagery obtained with SOFIA
on 2016 February 10.310 UT (mean UT). The images panels are 
(a) F077 = 7.70~\micron{}; FWHM = 5\farcs39, (b) F111 = 11.01~\micron{}; FWHM = 5\farcs23, 
(c) F197 = 19.70~\micron{}; FWHM = 4\farcs47, and 
(d) F315 = 31.36~\micron{}; FWHM = 4\farcs76. The vector 
indicating the direction of the comet's motion and the vector indicating the direction toward the Sun 
are also provided. Images were centroided and shifted using Fourier transform technique to the 
measured photocenter of the F197 image, and smoothed with a 3-pixel width median boxcar 
filter. \label{fig:fc-all-images}}
\end{figure*}

Images of comet C/2013 US$_{10}$ (Catalina) obtained during the 2016 February 09~UT flight 
are presented in Fig.~\ref{fig:fc-all-images}. Examination of the azimuthally averaged radial 
profiles of the comet in each filter reveals comet C/2103 US$_{10}$~(Catalina) exhibited extended 
emission beyond the point-spread function (PSF) of 
point sources observed with FORCAST under optimal telescope jitter performance in each 
filter.\footnote{\url{http://www.sofia.usra.edu/Science/ObserversHandbook/FORCAST.html}} 
Centroiding on the photocenter of the comet nucleus, photometry in an effective circular aperture
of radius 13 pixels, corresponding to 9\farcs984, with a background aperture annulus of inner 
radius 30 pixels (23\farcs58) and outer radius of 60 pixels (47\farcs16) was performed on the 
Level 3 pipeline co-added (*.COA) image data products
using the Aperture Photometry Tool \citep[APT v2.4.7;][]{2012PASP..124..737L}. 
The photometric aperture is $\simeq 3 \times$
the nominal point-source full width half maximum (FWHM), and encompassed the majority of
the emission of the comet and coma. Sky-annulus
median subtraction \citep[ATP Model B as described in][]{2012PASP..124..737L}
was used in the computation of the source intensity.
The stochastic source intensity uncertainty was computed using a depth of
coverage value equivalent to the number of co-added image frames.
The calibration factors (and associated uncertainties) applied to the
resultant aperture sums were included in the Level 3 
data distribution and were derived from the weighted average 
calibration observations of $\alpha$~Boo. 

The resultant SOFIA photometry is presented in Table~\ref{tab:simage_phot_tab}. 
For the SOFIA epoch of comet C/2013 US$_{10}$~Catalina, the coma did not appear to have jets or 
active areas creating discernible coma structures, by our visual examination of the photometric 
images divided by their azimuthally averaged radial profiles.

%
%


\begin{deluxetable*}{@{\extracolsep{0pt}}lccccc}
\tablenum{2}
%
%
\tablecaption{SOFIA Aperture Photometry and $\epsilon f \rho$ of Comet C/2013 US10 (Catalina)\label{tab:simage_phot_tab}}
\tablehead{
\colhead{Mean}\\
\colhead{Observation}\\
\colhead{Date} &  &\colhead{Fltr} &\colhead{Flux}\\
\colhead{UT 2016} &\colhead{InstCfg} &\colhead{$\lambda_{c}$} &\colhead{Density\tablenotemark{a}} &\colhead{$\lambda F_{\lambda}$} &\colhead{$\epsilon f \rho$}\\
\colhead{(dd-mm hr:min:s)} &\colhead{(Imaging)} &\colhead{($\mu$m)} &\colhead{(Jys)} &\colhead{($\times 10^{-16}$ W cm$^{-2}$)} &\colhead{(cm)}
}
\startdata
FORCAST (FOF276)\\
02-10T07:06:59.6 & SWC  & \phn7.70  & \phn3.238 $\pm$ 0.141 & 1.261 $\pm$ 0.055 & 7096 $\pm$ 308\\
02-10T07:29:47.6 & Dual & 11.01 &  \phn 13.8233 $\pm$ 0.415 & 3.767 $\pm$ 0.113 & 7881 $\pm$ 236\\
02-10T07:46:06.4 & Dual & 19.70 & 32.467 $\pm$ 0.423 & 4.941 $\pm$ 0.064        & 8522 $\pm$ 111\\
02-10T07:29:47.6 & Dual & 31.36 & 29.304 $\pm$ 0.303 & 2.801 $\pm$ 0.029        & 8677 $\pm$ \phn 90 \\[3pt]
FORCAST (FOF275)\\
02-09T08:09:02.3 & SWC  & \phn7.70  & \phn 3.133 $\pm$ 0.459 & 1.220 $\pm$ 0.179            &  6511 $\pm$ 954\\
02-09T08:31:48.8 & Dual & 11.01 & 17.961 $\pm$ 0.497 & 4.855 $\pm$ 0.135                    &  9721 $\pm$ 271\\
02-09T08:38:42.9\tablenotemark{b} & Dual & 11.01 & 16.128 $\pm$ 0.467 & 4.390 $\pm$ 0.127   &  8794 $\pm$ 255\\
02-09T08:46:50.0 & Dual & 19.70 & 40.100 $\pm$ 0.653 & 6.102 $\pm$ 0.099                    & 10159 $\pm$ 165\\
02-09T08:31:48.8 & Dual & 31.36 & 34.029 $\pm$ 0.809 & 3.259 $\pm$ 0.077                    &  9980 $\pm$ 232\\
02-09T08:38:42.9\tablenotemark{b} & Dual & 31.36 & 36.494 $\pm$ 0.575 & 3.489 $\pm$ 0.055   & 10470 $\pm$ 165\\
\enddata
\tablecomments{ }
\tablenotetext{a}{Measured in a circular aperture with a radius of 17\farcs664
centroided on the photocenter of the comet nucleus.}
\tablenotetext{b}{Aircraft climbing from 12.497 km to 13.106 km during observations.}
\end{deluxetable*}


\subsection{Dust Thermal Models of Infrared Spectra}\label{sec:dust_models}

Infrared spectroscopic observations are fitted with thermal models using standard 
spectral fitting techniques that minimize $\chi ^{2}$. Interpreting thermal models enables investigation 
into fundamental quantities of comet dust populations including: (1) bulk composition; (2) silicate 
structures of disordered (``amorphous silicates'') and/or crystalline forms (forsterite and enstatite);
(3) particle structures and size distributions; and (4) coma {\it bolometric} albedo. Refractory dust particles 
are much more robust in maintaining the chemical signatures from the time of 
formation \citep[see][]{2017RSPTA.37560260W}  than the highly volatile ices as well as semi-refractory 
organics with limited coma lifetimes \citep{2017RSPTA.37560260W, 2016Icar..278..301D}.  
Semi-refractory organics are known to exist through their limited lifetimes in comae, and are 
presumed to be organics in the dust that are modified while in the coma. These are the 
so-called `distributed sources', distributed to the coma by the dust particles. 
The semi-refractory organics are not (yet) observed in thermal IR spectroscopy but rather indirectly 
by the observed delayed release of molecules such as CO and/or H$_2$CO as described in
\citet{1999Natur.399..662D} and \citet{2008SSRv..138..179C} or by changes in the color of the 
scattered light \citep{2004A&A...424..325T}. Polarization properties of particles also are dependent 
upon organics \citep{2020P&SS..18304527H}.  \citet{2017RSPTA.37560260W} and 
\citet{2016Icar..278..301D} provide a detailed discussion of semi-refractory organics in cometary comae.

Thermal emission spectroscopy when combined with thermal modeling probes the particle composition from 
the optical active material in comet coma. A number of approaches have been employed to model the dust 
thermal emission and study the composition of cometary particles. Usually, these involve the simultaneous 
use of a number of different grain compositions (mineralogy), a size distribution, and a description of the 
particle porosity. Radiative equilibrium is assumed when deriving particle temperatures, which are 
strongly composition-dependent as well as particle-radii-dependent for low to moderate particle 
porosities. Particles of more highly absorbing compositions produce higher temperatures and higher flux 
density thermal emissions. To produce the combined emission of multiple compositions and integrated 
over grain size distributions, thermal models may employ an ensemble (sums) of individual particles 
of homogeneous dust materials \citep{2002ApJ...580..579H, 2011AJ....141...26H, 2017AJ....153...49H}, 
or may employ composition ``mixtures'' calculated using Effective Medium Theory  
\citep[see][]{2017MNRAS.469S.443B, 2017MNRAS.469S.842BErratum}. 

\begin{figure}[!ht]
\figurenum{5}
\centering
\includegraphics[trim=1.6cm 0cm 0cm 1.65cm,clip,width=1.10\columnwidth]{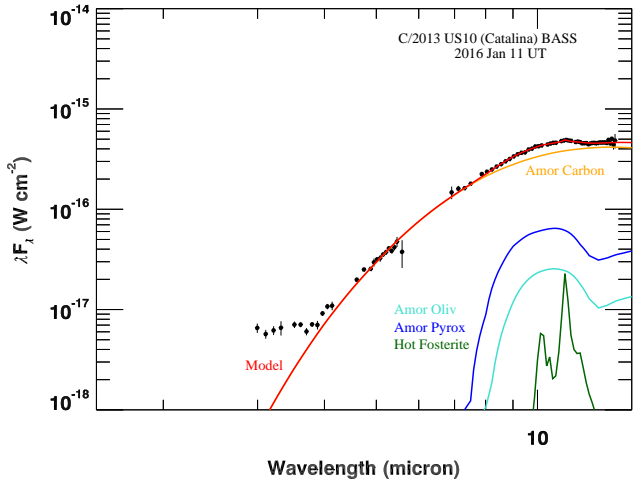}
\caption{Thermal model spectral energy distribution decomposition of comet 
C/2013 US$_{10}$ (Catalina) derived from the 3.0 to 14~\micron{} BASS spectrum 
obtained on 2016 Jan 10.61 UT at the NASA IRTF telescope. Gaps
in the spectra are due to regions of poor telluric transmission within
the continuous wavelength range covered by the instrument. The decomposition 
technique is used to determine the dust composition responsible for the 
observed coma emission at mid-infrared wavelengths. The solid red line is the best-fit 
model of the emission from the aggregate dust components, wherein the orange line 
represent the contribution from amorphous carbon, the dark blue solid line is the 
emission from amorphous pyroxene, the solid cyan turquoise  line depicts the 
amorphous olivine emission, and the green solid line depicts the crystalline olivine (``hot” forsterite).
The observed spectral data are the filled black circles with respective 
uncertainties.  The coma dust composition is dominated by amorphous carbon 
(dark material) and silicate grains with peak grain sizes (radii) of 0.5~\micron{} 
(Hanner grain size distribution). Some crystalline material is present.
\label{fig:dehsed_models_all}}
\end{figure}

\begin{figure}[!ht]
\figurenum{6}
\centering
\includegraphics[trim=1.6cm 0cm 0cm 1.65cm,clip,width=1.10\columnwidth]{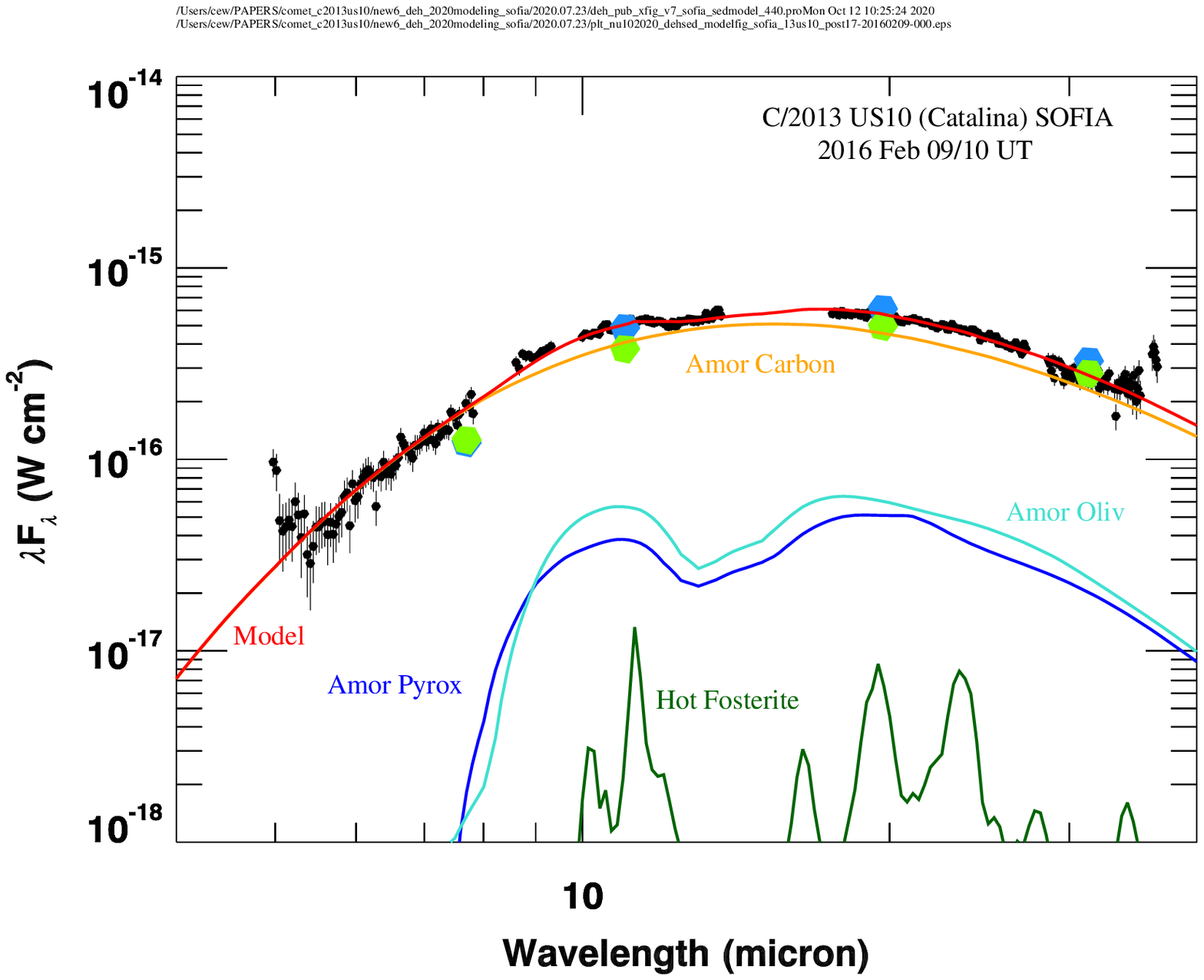}
\caption{Thermal model spectral energy distribution decomposition of comet 
C/2013 US$_{10}$ (Catalina) through all SOFIA grism segments, where the individual components
are the same as those described in Fig.~\ref{fig:dehsed_models_all}. 
The filled blue (2016 February 09 UT) and light green (2016 February 10 UT) circles 
superposed on the data points (black) are the photometry points taken from the 
FORCAST imagery in a circular aperture equivalent to the 
grism extraction area (the average for all grisms was 
17\farcs54 $\pm$ 0\farcs74, derived data product keyword PSFRAD) 
and are used to scale the spectral segments to the 
photometry. The coma dust composition is dominated by amorphous carbon 
(dark material) and silicate grains with peak grain sizes (radii) of 0.7~\micron{} 
(Hanner grain size distribution). Some crystalline material is present.
\label{fig:dehsed_allsofia}}
\end{figure}

At a given heliocentric ($r_{\rm h}$[au]) and geocentric ($\Delta$[au]) distance, the particle (dust) 
composition of the optically active grains, comprising a linear combination of discrete mineral 
components, porous amorphous materials, and solid crystals in a comet’s coma can be constrained 
by non-negative least-squares fitting of the thermal emission model spectra to the observed 
comet spectrum. The relative mass fractions and their respective correlated errors and the 
particle  properties including the porosity and size distribution, having invoked a Hanner 
grain-size distribution \citep[HGSD;][]{1983coex....2....1H} for $n(a)da$,  are given as a 
prescription for the composition of coma particles 
\citep[for details see][and references therein]{2018AJ....155..199H, 2011AJ....141...26H, 2002ApJ...580..579H}. 
The particle compositions of dust in the coma of comet C/2013 US$_{10}$~(Catalina) and 
relevant parameters from the best-fit thermal modeling are summarized in 
Table~\ref{tab:bf_sed_models_tab}. The uncertainties on the derived thermal model parameters
reflect the 95\% confidence limits that result from 1000 Monte Carlo trials \citep{2018AJ....155..199H}. 
Figures~\ref{fig:dehsed_models_all} and \ref{fig:dehsed_allsofia} show the resultant
models.

\begin{deluxetable*}{@{\extracolsep{0pt}}lcccc}
\tablenum{3}
%
%
%
\tablecaption{Derived Grain Composition of Comet C/2013 US10 (Catalina)\tablenotemark{a}
\label{tab:bf_sed_models_tab}}
\tabletypesize{\small}
\tablehead{
&\\[-8pt]
 &  \multicolumn2c{\underbar{BASS Spectra}\tablenotemark{b}} & \multicolumn2c{\underbar{SOFIA Spectra}\tablenotemark{c}}\\
 &                             & \colhead{Relative} && \colhead{Relative}\\
 &                             & \colhead{Mass} && \colhead{Mass}\\
 &                             & \colhead{Sub-\micron} && \colhead{Sub-\micron}\\
\multicolumn1l{Thermal Model SED Details} & \colhead{($N_{p}\times 10^{20}$)\,\tablenotemark{d}} & \colhead{Grains}
                 & \colhead{($N_{p}\times 10^{20}$)\,\tablenotemark{d}} & \colhead{Grains}
 }
\startdata
\underbar{Dust Components}\\[2pt]
Amorphous pyroxene (AP)& $0.132^{+ 0.014}_{- 0.014}$ & $0.242^{+ 0.022}_{- 0.023}$
&$0.630^{+  0.486}_{-  0.494}$ &$0.174^{+ 0.119}_{-  0.134}$\\
Amorphous olivine (AO)& $0.029^{+ 0.007}_{- 0.007}$ & $0.053^{+  0.014}_{- 0.013}$    
&$0.515^{+  0.342}_{-  0.342}$ &$0.142^{+ 0.115}_{-  0.098}$\\
Amorphous carbon (AC)& $0.567^{+ 0.003}_{- 0.003}$ & $0.473^{+ 0.017}_{- 0.015}$     
&$4.582^{+   0.116}_{-   0.115}$ &$0.574^{+ 0.083}_{-  0.075}$\\
Crystalline olivine (CO)& $0.072^{+ 0.009}_{- 0.009}$ & $0.232^{+ 0.022}_{- 0.024}$  
&$0.233^{+  0.327}_{-   0.233}$  &$0.111^{+ 0.123}_{-  0.111}$\\
Crystalline pyroxene (CP)& 0.000 \phn \phn \phn \phn & 0.000 \phn \phn \phn \phn
&0.000 \phn \phn \phn \phn & 0.000 \phn \phn \phn \phn\\[2pt]
\tableline
\underbar{Resultants}\\[2pt]
Total mass sub-\micron{} grains (gm) $\times 10^{8}$  & $0.942^{+   0.030}_{-  0.031}$ & \nodata
&$4.853^{+  0.745}_{-  0.609}$ & \nodata\\
Amorphous silicate dust fraction & $0.295^{+   0.015}_{-   0.014}$ & \nodata    
&$0.315^{+  0.066}_{-     0.067}$ & \nodata\\
Crystalline silicate dust fraction & $0.232^{+   0.022}_{-   0.024}$ & \nodata  
&$0.111^{+  0.123}_{-     0.111}$ & \nodata\\
Silicate to Carbon ratio\tablenotemark{$\dagger$}              & $1.116^{+   0.072}_{-   0.074}$ & \nodata
&$0.743^{+  0.264}_{-     0.220}$ & \nodata\\
Crystalline silicate mass to total silicate mass\tablenotemark{e} & $ 0.441^{+  0.032}_{-   0.035}$ & \nodata
&$0.260^{+  0.217}_{-   0.260}$ & \nodata\\
$a_{p}$(\micron)\tablenotemark{f} & 0.7  \phn \phn \phn \phn \phn \phn & \nodata
& 0.5  \phn \phn \phn \phn \phn \phn & \nodata \\
Fractal porosity ($D$)  & 2.727  \phn \phn \phn \phn & \nodata            
& 2.727  \phn \phn \phn \phn & \nodata \\[2pt]
\tableline
\underbar{Other Parameters}\\[2pt]
Hanner Grain-Size Distribution M : N &22.2 : 3.7  \phn \phn \phn \phn \phn \phn & \nodata
&13.6 : 3.4  \phn \phn \phn \phn \phn & \nodata \\
Reduced $\chi^2_\nu$ & $4.98$ \phn \phn \phn \phn \phn \phn & \nodata  
& $0.86$  \phn \phn \phn \phn \phn & \nodata \\
Degrees of freedom & 49 \phn \phn \phn \phn \phn \phn & \nodata    
& 168  \phn \phn \phn \phn & \nodata 
\enddata
\tablecomments{}
\tablenotetext{a}{Uncertainties represent the 95\% confidence level.}
\tablenotetext{b}{Comet on 2016 Jan 10 UT $r_{\rm h} = 1.30$ au, $\Delta = 0.76$ au.}
\tablenotetext{c}{Comet on 2016 Feb 09 UT $r_{\rm h} = 1.70$ au, $\Delta = 1.09$ au.}
\tablenotetext{d}{Number of grains, $N_{p}$, at the peak ($a_{p}$) of the Hanner grain size distribution (GSD).}
\tablenotetext{e}{$f_{cryst} \equiv m_{cryst}/[m_{amorphous} + m_{cryst}]$ where $m_{cryst}$ is the mass fraction of submicron crystals.}
\tablenotetext{f}{Peak grain size (radius) of the Hanner GSD.}
\tablenotetext{\dagger}{Ratio represents the bulk mass properties of the materials in the models.}
\end{deluxetable*}




\newpage
\subsubsection{Optical properties and IDP analogues}\label{sec:optical_properties_dehmodel}

A particle’s composition, structure (crystalline or amorphous), porosity, and effective radius ($a$) 
determine its absorption and emission efficiency, $Q_{\rm abs}$ (a grain’s absorption efficiency and 
emission efficiency are equivalent at any given wavelength by Kirchhoff’s Law). For an individual 
particle of effective radius $a$, 
$F_{\lambda}(a) \propto \pi \times a^2 \times Q_{\rm abs}(a)\times B_{\lambda}(T_{\rm dust}[a, \rm{composition}])$
where $B_{\lambda}$ is the Planck blackbody function, evaluated as a function of grain 
temperature, $T(\rm{K})$, particle size, and particle composition \citep{2002ApJ...580..579H}.

Our model uses optical constants ($n, k$) of five materials 
\citep[see][and references therein]{2002ApJ...580..579H, 2010LNP...815..203H, 2017RSPTA.37560260W}
 to compute $Q_{\rm abs}(a)$: a Mg-rich crystalline olivine, forsterite with a minerology of
 (Mg$_y$,Fe$_{(1-y)})_{2}$SiO$_{4}$, where $0.9\leq y \leq 1.0$ \citep{1998A&A...339..904J}; 
 a Mg-rich crystalline orthopyroxene, enstatite (MgSiO$_{3}$) \citep{1998A&A...339..904J}; amorphous 
 carbon \citep{Edoh_1983...PhdThesis},\footnote{Amorphous carbon is used by many modelers 
 \citep[see][]{2017MNRAS.469S.443B, 2017MNRAS.469S.598R}.}
and amorphous silicates of pyroxene-type and of olivine-type with compositions similar 
 to the stoichiometry of chondritic pyroxene (Mg$_{x}$,Fe$_{1-x}$)SiO$_{3}$ ($x = 0.5$\, i.e., Mg:Fe = 50:50)
and olivine (Mg$_{y}$,Fe$_{1-y}$)$_{2}$SiO$_{4}$ ($y = 0.5$\, i.e, Mg:Fe = 50:50) \citep{1995A&A...300..503D}. 
The amorphous silicates produce the broad width of the the 10~\micron{} silicate feature. When present, 
amorphous pyroxene generates a shorter wavelength shoulder on the 10~\micron{ } silicate feature. 
The crystalline materials are responsible for the sharp peaks in the IR spectra of comets at 11.15 to 11.2~\micron, 
19.5~\micron , 23.5~\micron, 27.5~\micron , and 33~\micron{} \citep[see][]{1997Sci...275.1904C, 2018AJ....155..199H}.
Absence of the latter crystalline spectral features in the observed IR SED does not imply that such 
species are absent in cometary comae \citep{2018AJ....155..199H}. However, without detection of spectral 
features, these species cannot be well-constrained by fitting thermal models to the mid-IR SEDs.
  
The optical properties of the materials used in the radiative equilibrium calculations for particle temperatures 
are derived from either laboratory-generated materials or mineral samples from nature.  Materials chosen 
for our thermal models have available optical constants and are found in or are analogous to materials in 
cometary samples. Crystalline silicates in Interplanetary dust particles \citep[IDPs,][]{2000Icar..143..126W}, 
\textit{Stardust}, UltraCarbonaceous Antarctic MicroMeteorites \citep[UCAMMs;][]{2010Sci...328..742D} 
are of olivine and pyroxene compositions with a range of Mg:Fe contents with typically 
1.0 $\leq$ y $\leq$ 0.5 and 1.0$\leq$ x $\leq$ 0.5 
\citep{2017RSPTA.37560260W, 2014GeCoA.142..240F, 2014GeCoA.144..277J, 2012GeCoA..76...68D, 2011Icar..212..896B}. 
Only Mg-rich crystalline olivine resonances have thus far been detected definitively 
in multiple comets using both the mid- and far-IR resonances.  Laboratory studies of crystalline olivine 
by \citet{2013ApJ...778...60K} show that with decreasing Mg-content (i.e., with $ y < 0.8$), the 
11.2~\micron{} peak shifts towards 11.4~\micron{} and the far-IR resonances dramatically change to 
different central wavelengths with different relative intensities. However, these more fayalitic crystalline 
olivine resonances have not been detected in comet comae.  

Amorphous silicates and amorphous carbon in thermal models are considered candidate ISM or dense 
cloud materials \citep{2017RSPTA.37560260W}. The outer cold disk where comet nuclei accreted is a likely 
reservoir of inherited interstellar grains \citep{2019SSRv..215...43S}. However, modeled characteristics of 
interstellar grains and measured cometary organics differ. \citet{2005A&A...433..979M} persistently suggest 
that the origin of the organic fraction of cometary IDPs is a different environment than the diffuse interstellar 
medium (DISM) because (a) the 3.4~\micron{} band of organics in anhydrous IDPs  is significantly narrower than
in the DISM (e.g., towards the Galactic Center that is a mixture of diffuse and dense cloud material) and 
(b) the aliphatic chains in IDPs are longer (less ramified) than in the DISM, based on the $-$CH$_2$/$-$CH$_3$ 
ratio in IDPs.

The Heterogeneous dust Evolution Model for Interstellar Solids (THEMIS) model 
\citep{2017A&A...602A..46J, 2016RSOS....360224J} predicts the formation and evolution of interstellar 
dust, from the harsher UV conditions of the ISM, through the DISM, the translucent clouds at the interface of 
and into dense clouds. In these regimes dust particles eventually either work their way out to less 
dense phases of the ISM and thus presumably are cycled into and out of phases, or the dust particles 
in the dense clouds make their way into protoplanetary disks such as our own. In translucent clouds 
THEMIS carbon-chemistry facilitates the growth of H-rich and aliphatic-rich matter, denoted a-C(:H), which 
accretes and then coagulates to tens of nm-size particles through a complex set of chemical reactions. 
The carbon-chemistry backbones are carbon belt-like molecules with aromatic bonds (n-cyclacenes) 
and an important process is the epoxylation of the surface materials. The carbonaceous particles, 
upon return to the harsh UV interstellar radiation field evolve ``towards an end-of-the-road H-poor and 
aromatic-rich a-C material'' \citep{2019A&A...627A..38J}. Carbonaceous matter in cometary samples 
appear significantly less dominated by aromatic moieties than implied by THEMIS models. 
\textit{Stardust} samples only reveal a small concentration of small PAHs \citep{2010M&PS...45..701C}. 
Carbon X-ray Absorption Near Edge Spectroscopy (C-XANES) spectra of \textit{Stardust} 
and IDP organics show saturated aliphatic carbon bonds are more recurrent than aromatic 
C\chemtwo{C} bonds as well as amorphous carbon being the only carbon form common between 
these samples and Bells, Tagish Lake, Orgueil and Murchison 
meteorites \citep{2009M&PS...44.1611W, 2017ApJ...848..113D}.  
Laboratory absorption spectra do not quantify amorphous carbon as it has no resonances, 
although its presence can be  discerned through Carbon X-ray Absorption Near Edge 
Spectroscopy \citep[C-XANES;][]{2004GeCoA..68.2577K, 2008LPI....39.2391M}.   

Amorphous carbon is found in many 
IDPs \citep{2004GeCoA..68.2577K, 2009E&PSL.288...44B, 2009M&PS...44.1611W, 2011Icar..212..896B} 
but amorphous carbon is not discussed for all IDPs \citep{2013EP&S...65.1159F, 2018PNAS..115.6608I} 
nor for all extraterrestrial particulate samples from primitive small bodies, specifically 
UCAMMs \citep{2018A&A...609A..65D, 2019A&A...622A.160M}. Despite a diversity of bonding 
structures in cometary organics \citep{2017MNRAS.469S.712B} as well as organic matter in asteroids, 
there is a severe paucity of optical constants \citep{2008ssbn.book..483D}. Typically, optical constants 
of relatively transparent tholens are combined with optical constants of the highly absorbing amorphous 
carbon to darken the models for surfaces of outer ice-rich bodies \citep{2008ssbn.book..483D}. Hence, 
amorphous carbon, which is devoid of aromatic bond IR resonances, is the best choice for the highly 
absorbing carbonaceous matter in models of dark surfaces of ice-rich bodies as well as for cometary 
coma particles. 

Amorphous silicates in thermal models are analogous to Glass with Embedded Metal and Sulfides (GEMS) in IDPs 
\citep[see][]{2006GeCoA..70.2371F,  2011Icar..212..896B, 2013GeCoA.107..336B, 2018PNAS..115.6608I, 2019LPI....50.2259S}. 
The ISM silicate absorption feature has spectral similarities to GEMS \citep{2013GeCoA.107..336B, 2019LPI....50.2259S} 
and radiation damage can explain the non-stiochiometry of GEMS \citep{2016ApJ...831...66J}. 
An alternative high-temperature formation scenario for GEMS is proposed for the 
protoplanetary disk \citep{2011GeCoA..75.5336K} but is challenged by discovery of GEMS with interior organic 
matter that could not have survived temperatures above 450~K \citep{2018PNAS..115.6608I}.  Amorphous 
silicates are a ubiquitous component of IR spectra of cometary comae and their radiation equilibrium 
temperatures require compositions of Mg:Fe$\approx$50:50 \citep{2002ApJ...580..579H}.

\subsubsection{The Hanner Grain Size Distribution (HGSD)}\label{sec:ss-hgsd}

Our modeling invokes the Hanner grain-size distribution $n(a) = (1 - {a_0} /a)^M~({a_0} /a)^N$ , 
where $a$ is the particle radius, where $a_0 = 0.1$~\micron \ is the minimum grain radius, 
and $M$ and $N$ are independent parameters \citep{1983coex....2....1H, 1994ApJ...425..274H}. The 
HGSD is a modified power law that rolls over at particle radii smaller than the {\it peak} radius 
$a_p = (M + N)/N$, which is constrained by the thermal model analyses.  

\subsubsection{Moderately porous particles}\label{sec:ss-mpps}

The optical properties of porous particles composed of amorphous materials may be calculated 
by incorporating ``vacuum'' as one of the material components \citep{1983asls.book.....B}. 
Porous grains are modeled with an increasing vacuous content as expected for hierarchical aggregation, 
using the porosity prescription or fractional filled volume given by $f = 1 - (a / 0.1~\micron)^{D - 3}$, where
$a$ is the effective particle radius, with the fractal dimension parameter $D$ ranging from $D = 3$ (solid) 
to $D = 2.5$ \citep[fractal porous but still spherical enough to be within the applicability of Mie theory 
computations;][and references therein]{2018AJ....155..199H, 2011AJ....141...26H, 2002ApJ...580..579H}.
Particle porosity affects the observed spectra of comets because the porous grains are cooler than solid 
grains of equivalent radius as their vacuous inclusions make them less absorbent at UV-visible 
wavelengths \citep{2002ApJ...580..579H}. The porosity prescription parameter $D$ is coupled 
with the grain size distribution slope parameter $N$, and the two parameters are simultaneously 
constrained when fitting IR SEDs. Increasing porosity (lowering $D$) decreases particle temperatures, 
which can be compensated for by increasing the relative numbers of smaller to larger grains by 
steepening the slope (increasing $N$) of the HGSD as illustrated in Fig.~2 of \citet{2002EM&P...89..247W}.  

An extremely porous particle  that is an aggregate of submicron compact monomers can have the 
same temperature as its monomers \citep[P(a)$_{\rm max} > $80\%; ][]{1997A&A...324..805X} or 
\citep[P(a)$_{\rm max} \geq $99\% with $a \geq$5~\micron; ][]{2007A&A...463.1189K}. However, IR 
spectra of comets are not well-fit by such extremely porous particles that are uniformly as hot as their 
submicron-radii monomers, regardless of particle size. Thermal models for observed IR spectra of comets 
need particle size distributions of moderately porous or solid particles.  For a comet near 1.5~AU, a 
HGSD has submicron- to micron-radii particles ($a_{peak} \leq 1$~\micron ) that produce the 
warmer thermal emission under the 10~\micron{} silicate feature and at shorter near-IR wavelengths,
as well as larger cooler particles producing the decline in the thermal emission at longer (far-IR) wavelengths. 
Compared to a size distribution of compact solid particles ($D = 3$), a size distribution of moderately porous 
particles (P$({\rm a)} \sim $66 to 86\%, $D = 2.727$ to $D = 2.5,$  a$_{\rm eff} = 5$~\micron) are cooler
and produce enhanced emission at longer wavelengths while still producing a silicate emission feature with 
the observed contrast compared to the local ``pseudo-continuum'' (see \S~\ref{sec:ss-10mss}). Hence, the 
thermal models constrain the porosity of the amorphous materials (amorphous silicates and
amorphous carbon), and the slope and the {\it peak} radius ($a_{peak}$) of the grain size 
distribution of \citep[see][]{2018AJ....155..199H, 2011AJ....141...26H, 2002ApJ...580..579H}. 

\subsubsection{CDE models of solid trirefringent silicate crystals}\label{sec:ss-cde}

Silicate crystalline particles are not well modeled as spheres by Mie Theory because of their anisotropic 
optical constants and irregular shapes \citep{2010ApJ...709..983K}. Crystals are not modeled as 
porous particles or as mixed-material particles using Effective Medium Theory because modeled 
resonant features do not match laboratory spectra of the same materials. Discrete solid crystals are 
better computed using the Continuous Distribution of Ellipsoids (CDE) approach 
\citep{2001A&A...378..228F} or the discrete dipole approximation \citep[DDA;][]{2013ApJ...766...54L}. 
Crystals of larger sizes than $\sim$1~\micron \ do not replicate the observed SEDs of 
comets \citep{2005Icar..179..158M}. CDE with c-axis elongated shapes reasonably reproduces laboratory spectra of 
crystalline forsterite powders \citep{2001A&A...378..228F} and serves as a starting point for our thermal 
models. Discrete solid crystals with sizes from 0.1 to 1~\micron{} are included in our admixture of coma 
dust materials. From our thermal models, we quote the relative mass fractions for the $\leq$1~\micron{}
portion of the HGSD in Table~\ref{tab:bf_sed_models_tab}.

\subsubsection{Comet Crystalline Silicates and Disk Transport}\label{sec:disk-xfer} 

The presence of crystalline silicate materials in cometary spectra and in cometary samples indicates 
transferal of materials that formed in the inner protoplanetary disk to the outer disk 
\citep{2017M&PS...52.1859W,2006Sci...314.1711B,2006Sci...314.1735Z} where volatile ices 
(H$_{2}$O, CO, CO$_{2}$) were extant along with dust particles to become incorporated into 
cometary nuclei \citep{2020SSRv..216..102R}. Crystalline silicates are relatively rare along 
lines-of-sight through the interstellar medium \citep[$\ltsimeq 5$\%,][]{2004ApJ...609..826K} and 
towards embedded young stellar objects or compact HII regions \citep[1 to 2\%, with a few sources 
at $>$3\%,][]{2020MNRAS.493.4463D}.  A significant crystalline silicate component in cometary dust 
has been clearly demonstrated by laboratory examinations of \textit{Stardust} \citep{2014GeCoA.142..240F} 
and IDPs \citep{2017M&PS...52..471B, 2009E&PSL.288...44B,1992Metic..27R.312Z}.
Crystalline silicate mass fractions (defined as 
$f_{crys} \equiv m_{cryst}/[m_{amorphous} + m_{cryst}]$ where $m_{cryst}$ is the
mass fraction of crystals) derived from thermal models of cometary IR SEDs typically 
are $\sim$20\% to 55\% \citep[][and Appendix I]{2011AJ....141..181W,  2018AJ....155..199H, 2011AJ....141...26H,
2007Icar..190..432H, 2002ApJ...580..579H, 2004ApJ...612L..77W}.
Detailed laboratory studies of cometary forsterite and enstatite crystals show a small fraction have 
mineralogical signatures of gas-phase condensation such as low iron manganese enriched 
(LIME) compositions \citep{2017M&PS...52.1612J, 2014GeCoA.142..240F}, 
$^{16}$O-enrichments commensurate with early disk 
processes \citep{2018LPI....49.2167D, 2016LPI....47.1584D,2017E&PSL.465..145D},
as well as condensation morphologies such as enstatite ribbons in anhydrous IDPs \citep{1999Sci...285.1716B}.

Moreover, \textit{Stardust} samples and some cluster IDPs contain olivine crystals with a wider range of 
Fe-contents (10\% $\ltsimeq$ Fe $\ltsimeq 60$\%) than the low Fe-contents of $\simeq$10\% to 20\% 
deduced from the wavelengths of the resonances of olivine crystals in cometary spectra 
\citep{2017RSPTA.37560260W, 1997Sci...275.1904C, 1996A&A...315L.385C}. 
It is a puzzle as to why the spectral signatures of Fe-bearing crystalline silicates are not spectrally 
detected in comets \citep{2017RSPTA.37560260W}.The Fe-bearing olivine 
crystals are analogous by their minor element compositions to olivine (Mg $\le$ 80\%) crystals in 
type-II chondrules and are called micro-chondrules or chondrule fragments 
\citep{2017M&PS...52..471B, 2014GeCoA.142..240F}.  In \textit{Stardust} samples, 
one 15~\micron-size type-II chondrule called ‘Iris’ has an age-date of $\geq$ 3 million-years 
(with respect to CAI formation) and is well-modeled as an isolated igneous 
system \citep{2015M&PS...50..976G}.

\textit{Stardust} samples pose a number of challenging questions for disk models about the 
formation of the nucleus of comet 81P/Wild~2. How did particles radially migrate as late as a 
few million years in disk evolution to the regime of volatile ices of H$_{2}$O, CO and CO$_{2}$? 
How did cometary dust minerals that condensed early in disk evolution persist in the disk long 
enough to be incorporated into this particular cometary nucleus, that is, persist and 
not be lost via the inward movement of particles? As of yet, satisfactory answers to either of 
these questions do not exist. 


Silicate crystals, specifically referring to forsterite and enstatite that are the abundant Mg-rich silicate 
crystalline species in comets and/or cometary samples \citep{2017RSPTA.37560260W}, condensed 
at temperatures near 1800~K or or possibly were annealed materials at temperatures near 
1100 to 1200~K in shocks \citep{2002ApJ...565L.109H} under low oxygen fugacity 
conditions \citep{2007prpl.conf..815W}. Radial transport may have occurred 
through a combination of protoplanetary disk processes including advection, diffusion, 
turbulence and aerodynamic sorting, meridional flows, disk winds, and/or planetary migration 
\citep{2019AJ....157..181V, 2011ApJ...740....9C,2010ApJ...719.1633H,2008arXiv0804.3377W,2004A&A...413..571G}.
Disk models with meridional flows \citep[see][]{2004A&A...413..571G} have been successful in 
predicting $\sim$20\% silicate crystalline mass fractions at disk radii of more than tens of AU in $<$1~million-years. 


Radial transport by advection can work through disk wind angular momentum transport 
\citep{2016ApJ...821...80B} but can also be produced by turbulent viscosity in the bulk of the disk. 
Radial transport by diffusion requires turbulence. It is generally thought that magento-hydrodynamical (MHD) 
turbulence occurs only in rarified upper layers of the disk atmosphere, if at all \citep{2016ApJ...821...80B}. 
However even without MHD effects, there are two recently-discussed hydrodynamical mechanisms for 
producing turbulence: convective over-stability (CO) and vertical shear instability 
(VSI) that are either individually or collectively operative in various locations
in the disk \citep[for example][]{2019ApJ...871..150P}. Meridional 2D flows are another robust 
feature of disk models when turbulence mechanisms are considered operative 
\citep{2019PASP..131g2001L, 2017A&A...599L...6S}. Yet, even the qualitative nature of this flow is 
debated. Meridional flows for 2D and alpha-disk models were outwards along the mid-plane and 
inwards above one scale height \citep[see][]{2004A&A...413..571G}. Recent 3D models of 
meridional flow show that the outward flow is above one scale height so particles that are 
lofted by turbulence to above one scale height above the mid-plane can move 
outwards \citep{2020arXiv200811195P, 2017A&A...599L...6S}. To date, meridional flows only are inferred from 
ALMA in $^{12}$CO observations of the $>$300 au outer disk regions of the $\sim$5 million-year old more 
massive Herbig Ae/Be system HD 163296 \citep{2019Natur.574..378T, 2019ApJ...878..116P}.
Large scale gas motions are not yet observed for analogs of our protoplanetary disk but cometary 
crystalline mass fractions suggest inner disk materials moved over large distances. 


Models without meridional flows also show outward movement of small particles, merely following the 
outward advective motion of the gas, at certain radii and times. \citet{2016ApJ...818..200E} 
show disk models (see their Fig.~15) with a range of dust particle masses in which the maximum 
disk radius reached by particles of a specific particle mass (i.e., size) increases with time, i.e., 
some particles do move outward and the smaller particles are more successful in moving 
outwards. Porous particles have larger aerodynamic cross sections compared to solid particles of 
the same mass so porous particles are favored in outward movement compared to solid 
particles \citep{2016LPI....47.2854E}. \citet{2012Sci...336..452C} simulate the 
migration of particles by randomized turbulent ‘kicks’, and thereby nicely illustrate the large-distance 
motions of some particles.


As a complement to transport within the disk, centrifugally driven disk winds may deposit 
particles with sizes $\ge1$~\micron{} to the outer disk at early times, which ``may be relevant to 
the origin of the 20~\micron{} CAI-like particle discovered in one of the samples returned from comet 
81P/Wild 2'' \citep{2019ApJ...882...33G}.  \citet{2019ApJ...887..156A} 
observed the brightest outburst to date from EX Lup using VLTI MIDI interferometry and VLT VISIR IR 
spectroscopy. Within five years practically all crystalline forsterite that had become enhanced in the inner 
disk disappeared from the surface of the inner disk. Over that time, the spectral resonances from olivine 
crystals shifted emphasis from the mid- to far-IR wavelengths indicating that the crystals experienced 
outward movement.


Disk models are challenged to effectively transport as well as maintain solids in the outer protoplanetary 
disk against the inward drift of particles, especially as particles grow to ‘pebble’ size and decouple 
from the gas. Models that treat particle coagulation as well 
as particle collisional destruction which maintain a population of fine-grained particles (i.e., smaller 
particles with lower Stokes numbers [$St_{\eta}$]) then outward movement of small particles 
with time occurs \citep[see][]{2016ApJ...818..200E}. Many studies have investigated how material 
that is injected into the disk spreads outwards and inwards with time 
\citep[for example,][]{2019PhDT.........5S}. When turbulence is a driving mechanism for radial
transport, then aerodynamics affects particle movements, and one can expect signatures of size 
sorting by $St_{\eta} \propto \bar\rho_{s} a$, where $a$ is the particle radius and
$\bar\rho_{s}$ is the average particle density \citep{2014Icar..232..176J, 2001ApJ...546..496C}.  \textit{Stardust} 
samples demonstrate that aerodynamic sorting in aggregate formation occurred for particles of olivine 
compared to FeS, which are denser than olivine \citep{2013ApJ...779..164W, 2012ApJ...760L..23W}. 
The \textit{Rosetta} mission’s imaging studies showed that comet 67P/Churyumov-Gerasimenko's 
particles are hierarchical aggregates of hundreds of microns to mm-size with components that are 
submicron to tens of micron in size \citep{2020P&SS..18204815L,2019A&A...630A..24G, 2016P&SS..133...63H}
\textit{Stardust} samples and \textit{Rosetta} particle studies are commensurate with 
the idea that aggregate particle components of submicron to tens of micron of size may be 
favored over larger solid particles in their outward movement to the disk regimes of 
comet-nuclei formation. 

\subsubsection{Revised specific density for Amorphous Carbon}\label{sec:ss-nurhoc}

Our thermal model adopts an amorphous carbon (Acar) specific density of $\rho_{\rm s}$(Acar) = 
1.5~g~cm$^{-3}$, from a quoted value of $\rho_{\rm s}$(Acar) = 1.47~g~cm$^{-3}$ \citep{1972JAP....43.3460W} 
measured for the same amorphous carbon material from which our optical constants were 
derived \citep{Edoh_1983...PhdThesis, 1994ApJ...425..274H}.\footnote{Edoh optical constants are of 
glassy carbon or of an amorphous carbon from the Plessey Company (U. K.) Ltd., Caswell, 
Towcester, Northants, England \citep{1972JAP....43.3460W}.}  This specific density of 
$\rho_{\rm s}$(Acar) used in these analyses of 
comet C/2013 US$_{10}$ (Catalina) herein represents a significant change from our prior thermal models 
and publications that used an assumed bulk density of carbon of 
2.5~g~cm$^{-3}$ \citep{1998ApJ...496..971L, 2002ApJ...580..579H}, which actually was a specific 
density slightly higher than that of graphite of 2.2~g~cm$^{-3}$ \citep{ROBERTSON2002129}. The 
relative mass fractions of carbonaceous matter and siliceous matter are important and allow us to 
take a detailed look at the carbonaceous contribution of comets to the hypothesized gradient of
 carbon in the solar system (\S\ref{sec:atomicCtoSi}) and as discussed by other authors 
\citep{2016M&PS...51..105H, 2017A&A...606A..16G, 2018A&A...609A..65D}.\footnote{If 
$\rho_{\rm s}$(Acar) = 2.5 ~cm$^{-3}$, then comet C/2013 US$_{10}$ (Catalina) would have 
yielded C/S $\approx 11$, which is greater than C/Si for any 67P/Churyumov-Gerasimenko
particle measured by COSIMA on \textit{Rosetta} \citep{2017MNRAS.469S.712B}. }

For completeness, in our thermal models the specific density of amorphous silicates 
is $\rho_{s}$(Asil) = 3.3~g~cm$^{-3}$ as discussed by \citet[][and references therein]{2002ApJ...580..579H}.

\subsection{Coma Dust Composition from Thermal Models}\label{sec:mir_analysis}

Comet C/2013 US$_{10}$ (Catalina) is a dynamically new (DN) Oort cloud with 
eccentricity of $\simeq 1.0003.$  Compositionally, the dust in the coma of comet C/2013 US$_{10}$ (Catalina)
is carbon-rich and this comet is among a subset of observed comets that are similarly carbon-rich, some 
of which are also DN. The carbon-rich dust particles of comet 67P/Churyumov-Gerasimenko were 
measured {\it in situ} to have by weight 55\% mineral and 45\% (carbonaceous) organic  
\citep[see Fig.~10, ][]{2017MNRAS.469S.712B}.  If we consider their mineral-to-organic ratio to be 
analogous to our silicate-to-carbon ratio then 67P/Churyumov-Gerasimenko has a ratio of 1.1 and 
C/2013 US$_{10}$ (Catalina) has ratios of 1.55 and 1.03 for 1.3 au (BASS) and 1.7 au (FORCAST), respectively.  
However, within the thermal model parameter uncertainties the silicate-to-carbon ratios are the same for 
both epochs. A decrease by a factor of 1.5 in the silicate-to-carbon ratio for the best-fit values 
between the two epochs is partly attributed to the definitive detection of crystalline forsterite at 1.3 au that 
increases the silicate mass fraction relative to the upper limit for forsterite at 1.7 au. Between the two epochs 
the amorphous carbon increases by a factor of 1.21 (see Table~\ref{tab:bf_sed_models_tab}). 

The dust particle population in comet C/2013 US$_{10}$ (Catalina) is characterized by a moderate particle 
porosity ($D = 2.727$).  Coma grains extend to submicron size particles, the HGSD (defined 
in \S~\ref{sec:ss-hgsd}) peaks at an average $a_{p}=\{0.7,0.5\}$~\micron, with a grain size distribution 
slope of $N = \{3.4, 3.7\}$, respectively, for the two epochs at 1.3~au and 1.7~au.  The derived
coma dust properties of C/2013 US$_{10}$ (Catalina) share similar characteristics with those found recently 
for some other long period Oort cloud comets, such as C/2007 N3 (Lulin) which is also DN 
\citep{2011AJ....141..181W}.

The HGSD slope of comet C/2013 US$_{10}$ (Catalina) is in the range of other comets, including Oort 
cloud comets, where typically $3.4 \leq N \leq 4$.  However, its HGSD slope is greater (steeper) than 
found for comet 67P/Churyumov-Gerasimenko, which has multiple measurements of its differential 
grain size distribution $n(a)da$ with slopes of $N = 3.0$ 
\citep[][]{2017MNRAS.469S.443B, 2017MNRAS.469S.842BErratum}, 
$N = 3.1$ \citep[][]{2017MNRAS.469S.598R}, $N = 3$ \citep[][]{2019A&A...630A..25D}, or
$N \simeq 2.7$ to 3.2 for $a < 100$~\micron{} and $N \simeq 1.8$ for 100$\leq a \leq$1000~\micron{}
\citep{2016A&A...596A..87M}. 

Examination of the SEDs of comet C/2013 US$_{10}$ (Catalina) obtained at two different epoch and the
thermal model derived parameters (Table~\ref{tab:bf_sed_models_tab}) enable us to deconstruct and 
decipher aspects of the inner coma dust environment (Figs.~\ref{fig:dehsed_models_all} and \ref{fig:dehsed_allsofia}). 
From the 58\% drop in the available ambient solar 
radiation between the 1.3~au (BASS epoch) and 1.7~au (SOFIA epoch) observations, one would expect
on average the particles on the coma to be cooler at the latter epoch. From the long wavelength 
shoulder ($\lambda \gtsimeq 12.5$~\micron) of the 10~\micron{} silicate feature and
longward, the SED measured at 1.7~au (Fig.~\ref{fig:obsflx_sofia_20160209}) shows enhanced emission at longer
wavelengths. Thus, the particles contributing to the far-IR emission are cooler at 1.7~au compared to 
those at 1.3~au as anticipated. However, the the thermal emission at 7.8~\micron{} and bluewards is
\textit{similar} for the two epochs. Hence, at 1.7~au the coma of comet C/2013 US$_{10}$ (Catalina) must
have an increased abundance of smaller warm amorphous carbon particles. Moreover, the number of 
dust particles in the coma at 1.7~au is increased over that at 1.3~au in order to produce about the same 
flux density of thermal emission at these two epochs with the cooler particles present at 1.7~au. 

There is evidence of a narrow 11.2~\micron{} silicate feature attributable to Mg-rich crystalline olivine 
\citep{2008SSRv..138...75W, 1994ApJ...425..274H}. This is borne out by the detailed 
thermal modeling of the SED which constrains the relative mass fraction of crystalline forsterite 
grains in the coma at 1.3~au.  The ratio of the crystalline silicate mass to the total  silicate mass was 
$\sim 0.44.$  The crystalline mass fraction determined for comet C/2013 US$_{10}$~(Catalina) is greater than 
that determined for other dynamically new comets such as C/2012 K1 (Pan-STARRS) studied with 
SOFIA \citep{2015ApJ...809..181W}. The derived values for each observational epoch are summarized 
in Table~\ref{tab:bf_sed_models_tab}.  

For the portion of the grain size distribution with radii $a \leq 1$~\micron{} (the submicron population), 
the silicate-to-carbon ratio is $1.116^{+0.072}_{-0.074}$ and $0.743^{+0.264}_{-0.220}$ 
at 1.3~au and 1.7~au, respectively (see Table~\ref{tab:bf_sed_models_tab}).  Compared to 1.7~au the higher 
silicate-to-carbon ratio  at 1.3~au is partly due to a factor of $\sim$1.25 less amorphous carbon combined 
with an increase in mass of silicates from the definitive detection of forsterite. This crystalline silicate material 
produces the sharp peak at 11.1 to 11.2~\micron{} \citep[][and references therein]{2010ApJ...709..983K} 
is relatively transparent outside of its resonances.  At 1.3~au, crystalline silicate mass fraction 
($f_{cryst}$) is 0.441$^{+0.033}_{-0.035}$ in the coma of comet C/2013 US$_{10}$ (Catalina) 
so forsterite crystals contribute significantly to the silicate-to-carbon ratio.  Crystalline silicates are 
tracers of radial migration of inner disk condensates or possibly shocked Mg-rich amorphous 
olivine so the 44\% crystalline mass fraction indicates significant radial transport of inner disk 
materials out to the comet-forming regime (see \S\ref{sec:disk-xfer}). 

\subsection{Silicate feature shape and strength}\label{sec:ss-10mss}

\begin{figure}[!ht]
\figurenum{7}
\centering
\includegraphics[trim=0.0cm 0cm 0cm 1.60cm,clip,width=0.95\columnwidth]{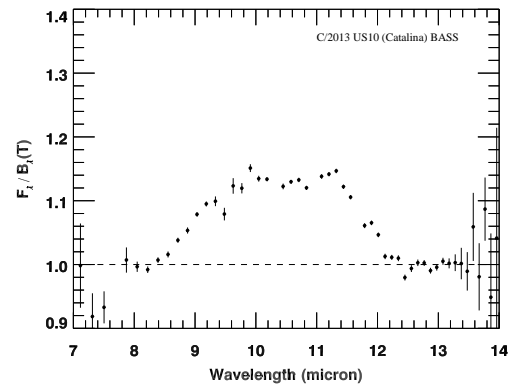}
\caption{Observed BASS flux density divided by a 265.3 K blackbody continuum derived from
the local 10~\micron{} `pseudo-continuum'  as defined by \citet{2004ApJ...612..576S} to highlight 
spectral details of the 10~\micron{} silicate feature.
\label{fig:s10_bass} }
\end{figure}

The spectral shape of the 10~\micron{} silicate feature can be revealed by dividing the observed flux 
by a local 10~\micron{} blackbody-fitted ‘pseudo-continuum.'  The shape of the 10~\micron{} silicate 
feature arises from emission from sub-micron- to at most several-micron-radii silicate particles in 
the the coma, depending on the porosity.  In thermal models, the ‘pseudo-continuum' has 
contributions from porous or solid amorphous carbon, which is featureless at all wavelengths. 
Thermal models require porous partciles ($D=2.7272$) for comet C/2013 US$_{10}$~(Catalina). 
Figure~\ref{fig:s10_bass} shows the silicate feature shape for comet C/2013 US$_{10}$ (Catalina) from the
BASS observations. The FORCAST mid-IR spectral data show a similar contrast silicate feature but with 
lower SNR as the BASS data, so these data are not included in the figure for clarity. 

The silicate strength parameter historically enables one to inter-compare the dust properties of 
different comets by quantifying the silicate feature contrast with respect to the local 
‘pseudo-continuum'  \citep{2004ApJ...612..576S, 2015ApJ...809..181W}. The 10~\micron{ } 
silicate feature strength, defined as $F_{10}/F_{c}$, where $F_{10}$ is the integrated silicate feature 
flux over a bandwidth of 10 to 11~\micron{} and $F_{c}$\, is that of the local blackbody 
`pseudo-continuum' at 10.5~\micron{} \citep{2004ApJ...612..576S}, is a metric that describes the 
contrast of silicate emission feature. We find the 10~\micron{} silicate feature to be weak in 
comet C/2013 US$_{10}$ (Catalina), approximately 12.8\% $\pm$ 0.1\% above the local 
`pseudo-continuum.'  The low silicate feature strength in comet C/2013 US$_{10}$ (Catalina) is similar 
to some other comets \citep{2004ApJ...612..576S, 2013EPSC....8..524S, 2015ApJ...809..181W, 2011AJ....141..181W}. 

A second metric used to compare dust properties of comets is the ratio of the SED color temperature 
(T$_{color}$) to the temperature that solid spheres would have at a given heliocentric distance ($r_h(\rm{au})$)
in radiative equilibrium with the solar insolation, 
$T_{\rm BB}(\rm K) = 1.1 \times 278\, (r_{\rm h})^{-0.5}$ \citep[see][]{1997EM&P...79..247H}. At 
the epoch of the the SOFIA observations, the \textit{combined}
grism 6.0 to 36.5~\micron{} SED can be fit with a single blackbody of temperature $239.5 \pm 0.5$~K, 
hence this ratio is $\simeq 1.02.$ The enhanced color temperature over a graybody, which is 
expected for the particles smaller than the wavelength,  often is historically referred to as 
``superheat'' $S$ \citep[see][]{1992Icar..100..162G}. The silicate strength parameter is somewhat 
correlated to $S$ \citep{2004ApJ...612..576S, 2015ApJ...809..181W}. For comet 
67P/Churyumov-Gerasimenko, $1.15\leq S \leq 1.2$, and $S$ is plotted along with the bolometric 
albedo at phase angle 90$^\circ$ (0.05 to 0.15) and the dust color (\% per 100 nm) 
\citep[][]{2019A&A...630A..22B}.  Comet C/2013 US$_{10}$ (Catalina) has a smaller value for $S$ than
comet 67P/Churyumov-Gerasimenko. 

C/2013 US$_{10}$ (Catalina) and 67P/Churyumov-Gerasimenko, both exhibiting a weak silicate feature
and are carbon-rich as determined from thermal modeling, provide a direct contradiction to 
older concepts commonly asserted in the literature. Commonly, many groups argued that some
comets totally lacked silicate features because their solid grains were radiating as graybodies and 
not displaying resonances because the grains were so large that the grains themselves were optically 
thick \citep{2005Sci...310..258A, 2005ApJ...625L.139L}. For comets with low dust production rates, 
estimation and subtraction of the nucleus' contribution to the SED is important. When combined with 
higher sensitivity observations and subtraction of the nucleus flux density, thermal models that 
integrate over a size distribution of particles with composition-dependent-dust-temperatures shows 
that the comets with comae particles whose HGSD has $a_p\leq 1$~\micron{} and that display 
weak silicate features are carbon-rich.  


\subsubsection{The ``Hot crystal model'' and SOFIA in the far-IR}\label{sec:ss-hotcrystals}

The SOFIA spectrum has enhanced emission that rises near 36~\micron{} but 
the observations do not extend to longer wavelengths to show a decline in flux density.  Laboratory 
absorption spectra of powders of pure-Mg forsterite show that the absorbance is about equal
at  33~\micron{} and 11.1~\micron{} \citep{2013ApJ...778...60K}, while the 19.5 and 23.5~\micron{}
features also having significant absorbance. The 33~\micron{} emission from pure-Mg 
forsterite (Fo100) is not detected in the far-IR. The slope of the HGSD is well constrained
by the SOFA data (given the low $\chi_{\nu}^{2}$). 

The SOFIA data provide important constraints on the crystalline resonances in the far-IR and 
on the slope of the HGSD (\S\ref{sec:ss-hgsd}). Our thermal models employ a ``hot crystal model" 
for the temperatures for forsterite and enstatite, where their radiative equilibrium temperatures of 
crystals are increased by a factor of 1.9$\pm$0.1 based on fitting the {\it ISO SWS} spectrum of 
comet C/1995 O1 (Hale-Bopp) \citep{2002ApJ...580..579H}. We speculate that hotter crystal 
temperatures may arise from crystals being in contact with other minerals that are more absorptive 
or from Fe metal inclusions such as 
``dusty olivines'' \citep[][]{1984LPSC...14..559K}, or ``relict'' grains \citep[][]{2017M&PS...52.1963R}. 

\subsection{Other mineral species not detected}\label{sec:ss-other-mins}

Within our SNR in the SOFIA mid- to far-IR SED, neither hydrated phyllosilicates that have 
far-IR resonances distinct from anhydrous amorphous olivine and amorphous pyroxene nor the 
very broad 23~\micron{} troilite (FeS,  submicron-sized) \cite[][]{2002Natur.417..148K} spectral 
signatures were seen \citep[see][]{2015Icar..260...60S}. Phyllosilicates, such as Montmorillonite, 
as well as carbonates have absorptions in the 5 to 8~\micron{} wavelength region 
\citep{1991Icar...94..191R, 2008Icar..195..938C} and neither of these compositions were detected 
in comet C/2013 US$_{10}$ (Catalina).


\subsection{The search for aliphatic and aromatic carbon}\label{sec:alamc}

The BASS spectrum spans the 3.0 to 3.5~\micron{} wavelength region where potentially 
the 3.28~\micron{} peripheral hydrogen stretch on a ring carbon macromolecule (PAH) and 
the 3.4~\micron{} -CH$_{2}$, -CH$_{3}$ aliphatic bonds arrangements that are prevalent in IDPs and 
\textit{Stardust} materials \citep{2013ApJ...765..145M} might be detectable. The analyses of a 
well-defined aliphatic carbon 3.4~\micron{} band on nucleus surface of 
67P/Churyumov-Gerasimenko is presented by \citet{2020NatAs...4..500R} and 
\citet{2017MNRAS.469S.598R} also argue for the presence for this feature
in coma observations.  The BASS spectrum spans the 3.0 to 3.5~\micron{} wavelength region where 
potentially the 3.28~\micron{} peripheral hydrogen stretch on a ring carbon macromolecule (PAH) and 
the 3.4~\micron{} -CH$_{2}$, -CH$_{3}$ aliphatic bonds arrangements that are prevalent in IDPs and 
\textit{Stardust} materials \citep{2013ApJ...765..145M} might be detectable. The analyses of a 
well-defined aliphatic carbon 3.4~\micron{} band on nucleus surface of 
67P/Churyumov-Gerasimenko is presented by \citet{2020NatAs...4..500R} and 
\citet{2017MNRAS.469S.598R} also argue for the presence for this feature in coma observations.
A broad 20\% deep 3.2~\micron{} features from organic ammonium salts also is discussed for
the nucleus \cite{2020Sci...367.7462P}. If the aliphatic material in comets is similar to that of IDPs then
laboratory absorption spectra by \citep{2005A&A...433..979M} of whole IDPs provide important information 
on the relative column densities of C atoms participating in different organic bonding groups including 
aliphatic bonds ($-$CH$_2$, $-$CH$_3$), aromatic (C\chemtwo{C}), carbonyl and carboxylic acid bonds 
in ketones, and ammonium salts.

\citet{2018ApJ...862L..16P} point to the possible presence of an organic emission feature near 
3.3~\micron{} in higher spectral resolution observations of comet C/2013 US$_{10}$ (Catalina) 
obtained on 2016 January 12 ($r_{h} = +1.3$~au) but do pursue any further detailed analyses.  
However, there are strong molecular ro-vibrational emission lines of C$_2$H$_6$ and CH$_3$OH 
in the 3.28 to 3.5~\micron{} region that significantly complicate deciphering underlying solid state organic 
features \citep{1995Icar..116...18B, 2006Icar..184..255D, 2009AJ....137.4538Y, 2017MNRAS.469S.443B}.  
Given these challenges, we do not report on detection of any aromatic or aliphatic features in the 
BASS data at our resolving power and sensitivity for comet C/2013 US$_{10}$ (Catalina). Thus, no 
spectral features were seen to indicate the presence of aromatic hydrocarbons (such as HACs, PAHs, 
a-C(:H) nano-particles) or aliphatic carbons in the coma of C/2013 US$_{10}$ (Catalina).

Comet C/2013 US$_{10}$ (Catalina) has one of the few reported 5 to 8~\micron{} 
wavelength spectrum from SOFIA (+FORCAST).  We searched for spectral signatures of vibration modes 
of C\chemtwo{C} bonds (6.25~\micron \, = 1600~cm$^{-1}$), based on a constrained search of the 
observed absorption features in laboratory studies of cometary-like polyaromatic organics in 
IDPs \citep{2005A&A...433..979M} and in the UCAMMs \citep{2018A&A...609A..65D} as well 
as asteroid insoluble organic materials \citep[IOM,][]{2017ChEG...77..227A}. The 
6.25~\micron{} C\chemtwo{C} resonances are not dependent on the degree hydrogenation 
or the number of peripheral hydrogen bonds compared to structural C\chemtwo{C} 
bonds \citep{2004GeCoA..68.2577K}.The UCAMMS are mass-dominated by organics, 
richer in N and poorer in O than with probable origins in the outer 
protoplanetary disk \citep{2009M&PS...44.1643D}.  We also searched for C\chemtwo{O} bonds 
(5.85~\micron \, = 1710~cm$^{-1}$).  There are tantalizing $\leq 3~\sigma$ fluctuations near 
1620~cm$^{-1}$ and 1510~cm$^{-1}$ that  are in the regions of C\chemtwo{C} stretching 
modes \citep[see Table~2 of][]{2014ApJ...780..174M}. However, the SNR is insufficient and the 
width of the fluctuations are narrow, narrower than the widths of the C\chemtwo{C} resonances in 
the UCAMMs that have a preponderance of organics such that their features dominate the 
5 to 8~\micron{} region. 

The lack of resonances from organics in the 5 to 8~\micron{} wavelength region does 
not discourage us from further searches in cometary comae for these bonding structures with the 
much higher sensitivity provided by the James Webb Space Telescope (JWST) and its instruments.

\subsection{Carbon and Dark Particles}\label{sec:carbon_sec}

We find amorphous carbon dominates the composition of grain materials in comet C/2013 US$_{10}$~(Catalina).
Dominance of carbon as a coma grain species was seen in other ecliptic comets including 
103P/Hartley~2\citep{2018AJ....155..199H} as well as the Oort cloud comets 
C/2007 N3~(Lulin) \citep{2011AJ....141..181W} and C/2001~HT50 
\citep{2006ApJ...651.1256K}. The outburst of dusty material from comet 
67P/Churyumov-Gerasimenko at 1.3~au was carbon-only-grains (with radii of order 0.1~\micron), 
as measured by VIRTIS-H \citep{2017MNRAS.469S.712B} and VIRTIS-M \citep{2018MNRAS.481.1235R} 
on \textit{Rosetta.} Comets can exhibit changes in their silicate-to-carbon ratio between observations epochs
and notably a few comets have had significant changes in their inner comae silicate-to-carbon ratios 
during a night's observations \citep[C/2001 Q4 (NEAT)][]{2004ApJ...612L..77W}, 
\citep[103P/Hartley~2][]{2018AJ....155..199H}, 
\citep[9P/Tempel~1][]{2007Icar..190..432H, 2005Sci...310..274S}. 

Our cometary comae dust atomic C/Si ratios are calculated using a number of suppositions and 
should be taken as indicative values. Cometary atomic C/Si ratios are of interest for comparison 
with \textit{in situ} studies of 67P/Churyumov-Gerasimenko  and 1P/Halley and of laboratory investigations 
of IDPs and UCAMMs.  The IDPs and UCAMMs are extraterrestrial materials likely to have originated from 
primitive bodies like comets and KBOs, 
respectively \citep[][and references therein]{2015PNAS..112.8965B, 2018A&A...609A..65D, 2019GeCoA.261..145B}.
We choose to compare C/Si of the submicron grain component determined from thermal models with 
bulk elemental composition measurements of IDPs (X-ray measurements).  We elect to not compare C/Si 
ratios derived from resonances (aliphatic 3.4~\micron, aromatic 6.2~\micron, and other bond in UCAMMs) 
because in laboratory base-line-corrected absorption spectra the amorphous carbon 
component would not be counted because it does not have a resonance. 

\subsubsection{Endemic Carbonaceous Matter in Comets}\label{sec:ss-endemic-c}

A dark refractory carbonaceous material darkens and reddens the surface of the nucleus of 
67P/Churyumov-Gerasimenko, the 
surface material also displays a 3.4~\micron{}  \citep{2020NatAs...4..500R} and a similar aliphatic 
feature is suggested to exist in the coma of 67P/Churyumov-Gerasimenko 
\citep{2017MNRAS.469S.598R}. We posit that the optical properties of amorphous carbon 
are representing well the dark refractory carbonaceous dust component observed in 
cometary comae through IR spectroscopy. Likely this dark refractory carbonaceous material 
is endemic to the comet's surface. Cosmic rays of a few 10 keV only damage a thin 
veneer of hundreds of nm of thickness \citep{2003CRPhy...4..791S, 2004Icar..170..214M, 2016Icar..272...32Q}.  
This damage effects the structure (amorphization) and the composition (destruction of C-H and O-H bonds 
by dehydrogenation) of the materials \citep{2004Icar..170..214M, 2015A&A...577A..41L, 2016Icar..272...32Q}. 
Typical particle radii on the nucleus surface of 67P/Churyumov-Gerasimenko is at least tens of microns 
based on the observed the red color of the surface at visible wavelengths \citep{2017P&SS..148....1J}, 
so cosmic rays do not damage the full particle volume. For example, IDPs studied by IR spectra 
indicate aliphatic bonds in particle interiors \citep{2005A&A...433..979M, 2013ApJ...765..145M, 2015EGUGA..17.2977F}  
but a lack of organic bonds in their near-surfaces possibly due to damaging ultraviolet light and particle 
radiation in space \citep{2004AdSpR..33...57F}.  Lastly, if the redeposition timescales for particles
lofted from the nucleus but not escaping its gravity are about 
the orbital period of comet 67P/Churyumov-Gerasimenko \citep{2020arXiv200513700M} then 
the ion-irradiation timescales on the surface, which have been shown to amorphize carbon 
bonds or damage silicates, are too short by orders of 
magnitude \citep{2004JRSp...35..487B, 2014Icar..237..278B, 2016Icar..272...32Q}.  

However, the surface properties of the DN comet like 
C/2103 US$_{10}$ (Catalina) may differ from the Jupiter-family comet like 67P/Churyumov-Gerasimenko.
A photon penetration depth of 1~\micron{} for cosmic-rays can induce chemical changes,  such as development 
of an organic crust due to the conversion of low molecular weight hydrocarbons into a web of bound
molecular species, from electronic ionization in dose time per 100eV per 16-amu (H$_2$O) in the
Local Interstellar Medium, which is a harsher environment than within the heliopause 
at $\sim$85~au \citep[see discussion in][]{2003CRPhy...4..791S}.
Comet C/2013 US$_{10}$ (Catalina) may have had a radiation damaged dust rime of up to a few cm 
depth, but DN comets can have their onset of activity at large heliocentric distances 
\citep{2009Icar..201..719M} where likely this material is shed when the comet's activity first 
turns on. Thus, the amorphous carbon is not from a radiation rime because of the insufficient volume 
of the nucleus that can be altered by radiation compared to the mass loss pre-perihelion.   
Coupled with the arguments about insufficient time scales for materials recently exposed on 
cometary surfaces from either erosion or re-deposition to be space weathered, we assert 
that the amorphous carbon that is in the observed comae of comet C/2013 US$_{10}$ (Catalina) 
is carbonaceous matter is endemic to the comet nucleus. Moreover, the fluence and time scales 
or temperatures that change carbon bonding structures typically are not reached in cometary 
comae. The material is refractory and stable. The dark refractory carbonaceous matter that is 
modeled with the optical constants of amorphous carbon (see \S~\ref{sec:optical_properties_dehmodel})
is endemic to comets. By the ubiquitous detection of a warm particle component in all cometary IR 
spectra observed to date, the carbonaceous matter is endemic to comets in general. 
  
If dark refractory carbonaceous matter is stable on the surface then this implies the matter 
will be stable in the coma, unless the temperatures are raised significantly. For example 
if the size distribution significantly changes to smaller sizes the latter would occur.  
Laboratory experiments demonstrate that amorphous carbon becomes graphitized 
at $\sim$3000~K \citep{2017ApJ...848..113D}.  Comae dust temperatures remain 
at $\ltsimeq400$~K dust compositions and 
particle sizes near 1~\micron-radii for comets near 1~au. The exception will be 
sun-grazers that come close or enter the solar corona. On the other hand, aliphatic carbon 
may survive temperatures as high as $\simeq 823$~K if associated with porous 
minerals \citep{2009M&PS...44.1611W}. In the outburst of 67P/Churyumov-Gerasimenko
at 1.3 au, comae dust temperatures reached 550 to 600~K and were modeled by tiny 
0.1~\micron-radii amorphous carbon particles 
\citep{2019A&A...630A..22B, 2017MNRAS.469S.842BErratum, 2018MNRAS.481.1235R}.  
Thus, comet comae dust particles do not reach such high temperatures as $\simeq 823$~K
to  destroy aliphatic carbon when comets are near 1 au. 

The contribution of amorphous carbon is variable between comets. In some comets, the 
contribution of amorphous carbon is temporally variable:~103P/Hartley~2 \citep{2018AJ....155..199H}, 
C/2001 Q4 (NEAT) \citep{2004ApJ...612L..77W}, and the after the kinetic-impactor encounter 
in inner coma of 9P/Tempel~2 \citep{2005Sci...310..274S, 2007Icar..190..432H}. The variability of 
amorphous carbon between comets and the temporally variability for a few comets gives clues 
to the diversity of protoplanetary disk reservoirs out of which comet nuclei formed. The variability in 
silicate-to-amorphous carbon ratios for an individual comet also may be related to the 
size sales of variable-compositions of the nucleus \citep{2007Icar..187..332B}, to 
jets \citep{2004ApJ...612L..77W}, or variations coupled to changes in solar insolation in 
different parts of comets orbits \citep[seasonal effects;][]{2020Icar..33513421C}. These 
variations asserted for the nucleus are tied to the hypothesis that the refractory dust 
particle compositions observed in the coma are endemic to the comet. 

\subsection{Amorphous carbon and other forms of carbon}\label{sec:ss-ac-and-otherforms}

Amorphous carbon is the one carbon bonding structure common to IDPs, Stardust, and 
four carbonaceous chondrites including Bells, Tagish Lake, Orgueil, and Murchison 
\citep{2009M&PS...44.1611W, 2017ApJ...848..113D}. The amorphous carbon bonding structure 
is observed specifically through C-XANES \citep{2008GeCAS..72R.603M} and in \textit{Stardust} 
particles from comet 81P/Wild~2 \citep{2008M&PS...43..315M}. In addition to C-XANES spectra, 
regions of some IDPs are described a poorly graphitized or highly disordered 
carbon \citep{1993Metic..28R.448T, 1993GeCoA..57.1551T}. 

Other organic bonding structures besides amorphous carbon that are found in cometary 
samples (IDPs and \textit{Stardust}) are: aliphatic, aromatic, and rarely graphitic. IDP organic 
matter generally occurs as aliphatic-dominated rims \citep{2008EM&P..102..447F, 2015EGUGA..17.2977F}, 
rims on mineral grains with aromatic (C\chemtwo{C}) and carbonyl group (C\chemtwo{O}) 
bonds \cite{2013EP&S...65.1159F}, (non-graphitized) aliphatic or aromatic macromolecular 
material \citep{2017ApJ...848..113D} as submicron-sized pieces associated with mineral 
crystals \citep{2009M&PS...44.1611W}, or as a matrix \citep{2014Icar..237..278B}. 
In one IDP, different bonding structures of carbon occurs in micron-sized regions and where 
amorphous carbon was mixed with GEMS \citep{2014Icar..237..278B}. Two IDPs show 
N-rich organic rims on GEMS that are in turn are inside other GEMS, indicating two formation 
epochs, and their specific organic matter requires particle temperatures remained cooler 
than $\sim$450~K \citep{2018PNAS..115.6608I}. Cometary carbonaceous matter 
is sometimes referred to as polyaromatic when there are significant moities of 
aromatic C\chemtwo{C} bonds. UCAMMs are noted for abundant aromatic material as 
well as for their N\chemtwo{C} and N\chemone{C} bonds 
\citep{2018A&A...609A..65D, 2019A&A...622A.160M}. 

Only four cometary samples display graphitic carbon bonding structures as witnessed through
C-XANES. Two of  these are from \textit{Stardust} samples, seen as halos on Fe grain cores 
which are hypothesized to have formed at high temperatures and at low oxygen fugacity in 
the protoplanetary disk \citep{2017ApJ...848..113D}, and in two IDPs (L2021C5, L2021Q3)
where its close proximity to other bonding structures is discussed respectively
by \citet[][]{2014Icar..237..278B} and \citep[L2021Q3][]{2016A&A...596A..87M}.  Graphite can 
be formed at high temperatures ($\gtsimeq 3273$~K) although there are lower temperature 
processes that form graphite \citep{2009M&PS...44.1611W}.  Ion bombardment of amorphous 
carbon is a competing process between amorphization and graphitization and 
this process depends on the structure of the starting amorphous carbon \citep{2011Icar..212..896B}.  
Raman spectroscopy of one IDP shows ``localized micrometer-scale distributions of extremely 
disordered and ordered carbons'' \citep{2011Icar..212..896B}.  

In summary, cometary carbonaceous matter is macromolecular \citep{2017ApJ...848..113D} 
and not strictly aromatic (containing aromatic bonds) like meteoritic 
IOM \citep{2007GeCoA..71.4380A}, as well as highly variable in composition and structure. 

\subsection{Cometary comae elemental C/Si ratios}\label{sec:atomicCtoSi}

In the following discussion, we investigate the plausible implications of cometary comae thermal 
model’s relative mass fractions (i.e., the mass fraction of amorphous carbon to 
the mass fractions of the amorphous and crystalline silicates) on the elemental 
abundance ratio of C/Si.  We compare inferred elemental ratio C/Si for
comet C/2013 US$_{10}$ (Catalina) from thermal models to the C/Si ratio determined for 
IDPs using Scanning Electronic Microscopy with Energy Dispersive
X-ray analysis \citep[the SEM-EDX method,][]{1993Metic..28R.448T}, and by mass 
spectrometry for comet 1P/Halley, and comet 67P/Churyumov-Gerasimenko (COSIMA).  

We will show that the relative mass fractions of C/Si derived from our thermal models of 
comet C/2013 US$_{10}$ (Catalina) and a handful of other recently observed and modeled 
comets are consistent with  the average C/Si $= 5.5^{+1.4}_{-1.2}$ derived by COSMIA for thirty 
67P/Churyumov-Gerasimenko particles \citep{2017MNRAS.469S.712B}, for 1P/Halley particles 
measured by Vega-1 and Vega-2 mass spectrometers during spacecraft encounters, 
and also for the upper range of C/Si for IDPs \cite[see][]{2015PNAS..112.8965B}.  The enigmatic 
comet C/1995 O1 (Hale-Bopp) with is propensity of submicron crystalline 
silicates \citep{2002ApJ...580..579H} also is included in our analysis to demonstrate its lower
C/Si ratio that is in the lower range of the IDP C/Si ratios \citep{2017MNRAS.469S.712B} and 
also close to the range determined for CI chondrites \citep{2015PNAS..112.8965B}.

Our cometary comae dust C/Si atomic ratios are calculated using a few suppositions and 
should be taken as indicative values, which are of interest for comparison with {\it in situ} studies of 
67P/Churyumov-Gerasimenko and 1P/Halley and of laboratory investigations of IDPs and 
UCAMMs \citep{2005A&A...433..979M, 2014Icar..237..278B, 2017MNRAS.469S.712B, 2018A&A...609A..65D}. 
The IDPs and UCAMMs are extraterrestrial 
materials likely to have originated from primitive bodies like comets and KBOs, 
respectively \citep[][and references therein]{2009M&PS...44.1643D}. 
Unlike laboratory measurements of IDPs, micrometeoritic samples, or \textit{Stardust} 
particles which generally are the measure of single grains or isolated domains 
within a matrix, values returned from remote-sensing spectroscopic observations 
represent a coma-wide measure from a large ensemble of thermally radiating dust 
particles of various radii. 

Our suppositions in deriving C/Si atomic ratios are: (a) amorphous 
carbon is a good optical analog for dark highly absorbing carbonaceous matter in cometary 
comae and (b) thermal model relative mass fractions derived for amorphous carbon 
are representing a significant fraction of the carbonaceous matter in the coma 
\S{\ref{sec:counting_carbon}}. 

\subsubsection{Counting Carbon Atoms}\label{sec:counting_carbon}

We are comparing the C/Si atomic ratio derived for cometary samples using different techniques.
Mass spectroscopy directly measures the elemental C/Si ratio, which is the method for 
{\it in situ} measurements. However, non-destructive techniques that allow counting the carbon 
atoms in IDPs or \textit{Stardust} samples depend on the method. X-ray SEM-EDX techniques 
\citep{1993Metic..28R.448T} can count all the carbon atoms whereas IR absorption spectroscopy 
counts the carbon atoms involved in the observed resonances. Laboratory IR absorption 
spectroscopy measures the C/Si by converting the integrated band strengths into the number of 
atoms for aliphatic and/or aromatic bands compared to the 10~\micron{} silicate band 
\citep{2005A&A...433..979M, 2014Icar..237..278B}. Laboratory absorbance spectroscopy 
fits and subtracts a spline baseline to yield a linear baseline for the purpose of integrating 
the observed band strengths \citep[see][]{2005A&A...433..979M}. Amorphous carbon is not observed 
in absorbance in spectroscopy of IDPs because amorphous carbon lacks spectral resonances. 
To make a comparison between cometary C/Si derived from thermal models of amorphous carbon 
and C/Si derived from laboratory measurements and {\it in situ} measurements, 
we choose to employ the SEM-EDX measurements that are counting the carbon atoms but not 
discerning the carbon bonding structures. 

Currently we cannot claim knowledge of aliphatic and aromatic content in comet 
comae dust populations of multiple comets via IR spectroscopy. If we cannot detect signatures of 
these bonding structures, we cannot definitely determine their contribution to the observed 
emission. However, we can use IDPs to indicate what the potential increase in C/Si might be if 
the aliphatic or aromatic bonds were spectroscopically detected.  

We can examine what C/Si atomic ratios are derived from organic features in laboratory 
absorbance spectra of IDPs and compare to the C/Si derived for comets using thermal modeling 
of the warm particle component that is modeled with amorphous carbon. Many IDPs show the 
aliphatic 3.4~\micron{} feature. The 3.4~\micron{} feature is composed of the aliphatic 
CH$_{2}$ symmetric vibration (at $\sim$2850~cm$^{-1}$), the CH$_{2}$ asymmetric 
vibration (at $\sim$ 2922~cm$^{-1}$) and the weaker CH$_{3}$ asymmetric vibration 
(at $\sim$ 2958~cm$^{-1}$) as discussed in \citet{2005A&A...433..979M}. In six IDPs, 
the 3.4~\micron{} aliphatic carbon features yield 0.27~$\leq$~C/Si~$\leq$~1.4 with a mean 
C/Si $= 0.55\pm 0.43$ \citep[see Table~4 of ][]{2005A&A...433..979M}. For three out of the six 
IDPs, acid dissolution of the silicates allowed the detection of the intrinsically weaker aromatic 
skeletal ring stretch C\chemtwo{C} at 6.25~\micron \ (1600~cm$^{-1}$),  which raises the atomic 
ratios for these three IDPs from C$_{\rm aliphatic}$/Si $= \{0.78, 0.11, 0.55\}$ to 
C$_{\rm aliphatic + aromatic}$/Si $= \{19.4, 3.1, 5.1\}$ \citep[see Table~5 of ][]{2005A&A...433..979M}.  

Most IDPs, however, do not possess an aromatic 3.28~\micron{} feature from C-H peripheral bonds 
on C\chemtwo{C} skeletal rings.  \citet{2004GeCoA..68.2577K} suggest the lack of the 
3.28~\micron{} aromatic feature is due to ``much of the carbonaceous matter
is comprised very poorly graphitized carbon, possessing only short range order 
($<$2 nm), or very large PAH molecules.'' The  C\chemtwo{C} bonds that are better tracers 
of the aromatics than the peripheral C-H bonds. As yet, no comet has been observed with 
organic features that are of comparable absorbance as the silicate features as observed in 
absorption spectra of three UCAMMs, where organic absorbances are as strong as for the 
silicate features \citep{2018A&A...609A..65D}. As other authors suggest, we 
infer comets have less ``outer disk processed organics'' than UCAMMs. This conjecture is
also supported by noting the ratio of nitrogen-to-carbon (N/C) in 
67P/Churyumov-Gerasimenko is less than the N/C in UCAMMs 
\citep{2017MNRAS.469S.712B, 2018A&A...609A..65D}. If IR spectra of cometary comae 
were to detect the 3.4~\micron{} feature at about the same contrast to the silicate feature 
as is in laboratory absorbance spectra of IDPs 
\citep{2005A&A...433..979M, 2014Icar..237..278B, 2016A&A...596A..87M}, 
then we may infer that C/Si for our comets that we analyze might increase $\sim$20\%.

\subsubsection{The C/Si gradient in the Solar System}\label{sec:ss-c2sgradient}

\begin{figure}[!b]
\figurenum{8}
\centering
\includegraphics[trim=2.0cm 2.0cm 0cm 0.5cm,clip,width=1.10\columnwidth]{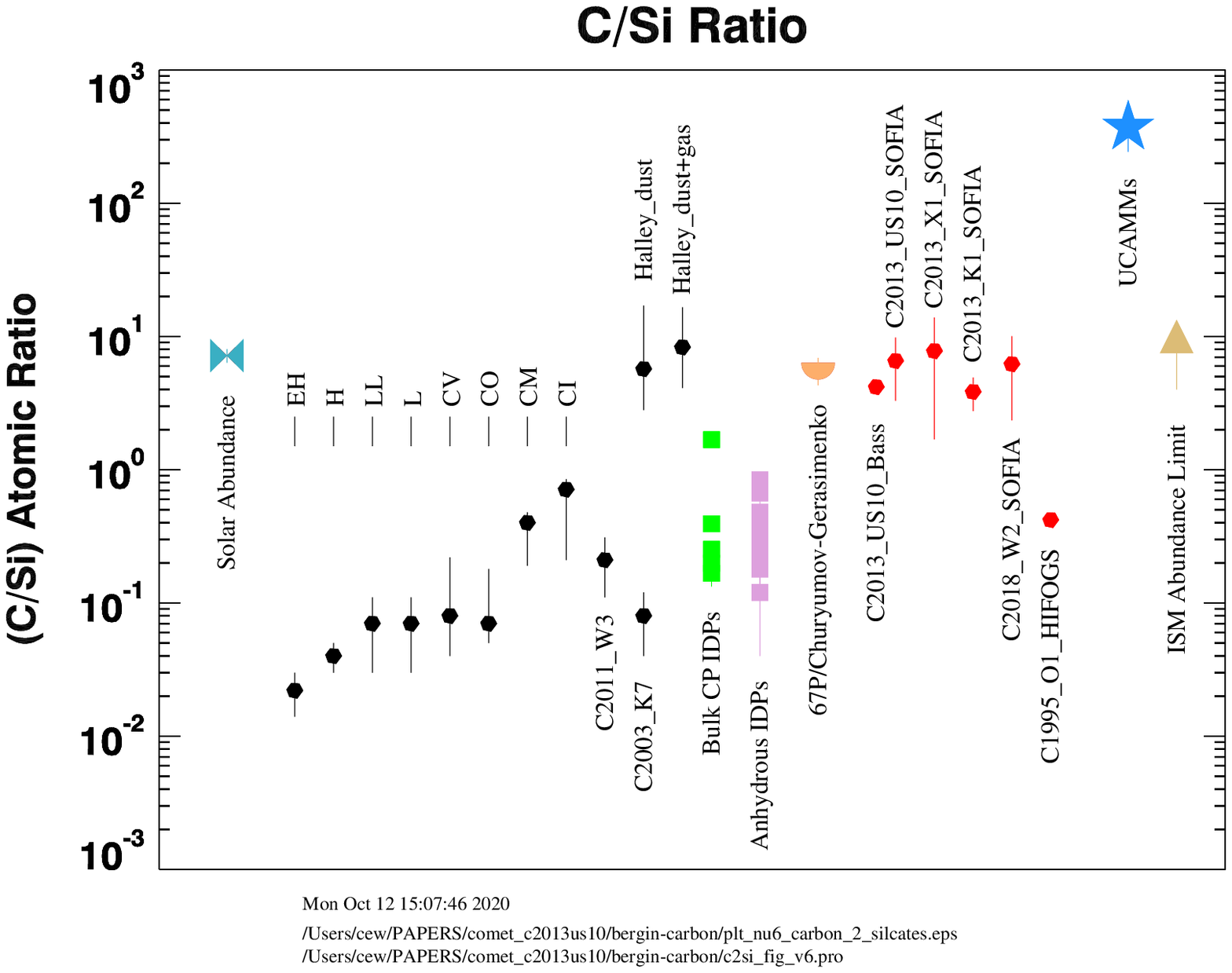}
\caption{Atomic carbon-to-silicon ratio for various small bodies relevant to
comets. Ratios from \citep[][supplemental appendix]{2015PNAS..112.8965B}
are given with black filled circles, while those derived for comets
from our thermal model analysis (see \S\ref{sec:ss-c2sgradient}) are 
indicated with the red filled circles. Also depicted by the filled green squares are
five values for chondritic porous (CP) interplanetary dust particles (IDPs) from
\citet{2004GeCoA..68.2577K}, the purple filled squares are values for 19
anhydrous IDPS measured by \citet{1993Metic..28R.448T}, and the half-filled circle
is average value of the Ci/S atomic ratio of comet 67P/Churyumov-Gerasimenko
particles studied by \citet{2017MNRAS.469S.712B}. The blue star denotes the
values for the UltraCarbonaceous Antarctic MicroMeteorites (UCAMMs) while 
the limit to the interstellar medium C/Si atomic ratio, brown triangle, is from \citet{2018A&A...609A..65D}. 
Both C/2011 W3 and C/2003 K7 are sun grazing comets and the determination of the
C/Si atomic ratio in these objects is derived from ultra-violet measurements when these comets 
were in the solar corona \citep[see][and references therein]{2015PNAS..112.8965B}.
The solar abundance values are taken from \citet{2010ASSP...16..379L} and references
therein.
\label{fig:c2s_plot}}
\end{figure}

We derived the C/Si atomic ratio using the thermal model dust 
compositions (and relevant atomic amu) described in \S{\ref{sec:mir_analysis}} and 
the relative masses of the submicron grains for each composition returned 
from the best-fit thermal model. The asymmetric uncertainties 
in the relative masses derived from the thermal models were `symmetrized' following 
the description discussed by \citet[Method\#2,][]{2017ChPhC..41c0001A}, cognizant of
the limitations to this approach \citep[see][]{2019Metro..56d5009P,2003sppp.conf..250B} to 
enable standard error propagation techniques. The carbon to silicon atomic ratio
is defined as:

\begin{equation}
\centering
\begin{split}
\frac{C}{Si} &= \frac{C}{\Sigma(Si^{dust}_{species})}\\
&= \frac{N_{p}(C) \cdot C_{amu}}{\frac{N_{p}(AO)}{\alpha} + \frac{N_{p}(AP)}{\beta} + \frac{N_{p}(CO)}{\delta} + \frac{N_{p}(CP)}{\gamma}}
\end{split}
\end{equation} 

\noindent where

\begin{equation}
\begin{split}
\alpha &=  \frac{(0.5 \cdot  \rm{Mg}_{amu} + 0.5 \cdot \rm{Fe}_{amu}) \times 2 + \rm{Si}_{amu} + 4 \cdot O_{amu}}{\rm{Si_{amu}}} \\
\beta   &= \frac{(0.5 \cdot  \rm{Mg}_{amu} + 0.5 \cdot \rm{Fe}_{amu})  + \rm{Si}_{amu} + 3 \cdot O_{amu}}{\rm{Si_{amu}}} \\
\delta   &= \frac{(\rm{Mg}_{amu}) \times 2 + \rm{Si}_{amu} + 4 \cdot O_{amu}}{\rm{Si_{amu}}} \\
\gamma   &= \frac{(\rm{Mg}_{amu})  + \rm{Si}_{amu} + 3 \cdot O_{amu}}{\rm{Si_{amu}}} \\
\end{split}
\end{equation} 

\noindent are the $\alpha, \beta, \gamma, \, \rm{and} \, \delta$ are the number of Si atoms per unit mass, 
and the values for $N_{p}$ (the number of grains at the peak [$a_{p}$] of the HGSD) 
are found in Table~\ref{tab:bf_sed_models_tab}.  Table~\ref{tab:ac2si_tab} summarizes
derived the Ci/Si atomic ratios for comet C/2013 US$_{10}$ (Catalina) and other 
comets observed with SOFIA (+FORCAST) as well as comet C/1995 O1 (Hale-Bopp)
\citep{2002ApJ...580..579H}. The C/Si atomic ratio for the comets in Table~\ref{tab:ac2si_tab}, 
UCAMMs \citep[data from][]{2018A&A...609A..65D}, and IDPs and other comets
\citep[data from][]{2015PNAS..112.8965B} are presented in Fig.~\ref{fig:c2s_plot}. 
Recent measurements of solar cosmic abundances creates an upper limit for the ISM C/Si of
10 as discussed in \citet[][and references therein]{2018A&A...609A..65D}.
UCAMMs are above the solar cosmic abundance limit. Thus those who study UCAMMs suggest 
that their organics have sequestered carbon from the gas phase and converted it to a solid phase 
in the cold outer disk or on the surfaces of nitrogen-rich cold body surfaces because of their 
enhanced N/C ratios \citep{2013Icar..224..243D, 2018A&A...609A..65D}.  As measured or 
computed, cometary comae appear to lack the high C/Si ratios of UCAMMs.  


%



\begin{deluxetable*}{@{\extracolsep{0pt}}lccc}

\tablenum{4}
%
%
%
%
\tablecaption{Derived Carbon-to-Silicon Atomic Ratio of Comets\tablenotemark{a}
\label{tab:ac2si_tab}}
\tablehead{
\colhead{Comet} &\colhead{Telescope/Instrument} &\colhead{C/Si}  &\colhead{(Ref.)}
}
\startdata
C/2013 US10 (Catalina)    & IRTF (+BASS)      & $4.180 \pm 0.308$  & (1)\\
C/2013 US10 (Catalina)    & SOFIA (+FORCAST)  & $6.556 \pm 3.262$  & (1)\\
C/2013 K1 (Pan-STARRS)    & SOFIA (+FORCAST)  & $3.841 \pm 1.086$  & (2)\\
C/2013 X1 (Pan-STARRS)    & SOFIA (+FORCAST)  & $7.781 \pm 6.091$  & (3)\\ 
C/2018 W2 (Africano)      & SOFIA (+FORCAST)  & $6.204 \pm 3.858$  & (3)\\
C/1995 O1 (Hale-Bopp)     & IRTF (+HIFOGS)    & $0.420 \pm 0.001$  & (4)\\
\enddata
\tablecomments{References: (1) This work. (2) \citet{2015ApJ...809..181W}. (3) \citet{Woodward2020...inprep}. (4) \citet{2002ApJ...580..579H}. }
\tablenotetext{a}{Computed from relative mass ratios of thermal model dust compositioni components, adopting
an amorphous carbon density $\rho = 1.5$~g~cm$^{-3}$. Appendix A provides the complete model tables, including
revision to C/2013 K1 (Pan-STARRS) and C/1995 O1 (Hale-Bopp) models resulting from an amorphous carbon specific density $\rho_{s}(\rm{ACar}) = 1.5$~g~cm$^{-3}$.}
\end{deluxetable*}


Comets by their C/Si appear to be sampling similar abundances of carbon in the 
optically active composition of comae particles as SEM-EDX-derived C/Si ratios are measuring 
for IDPs. Many but not all comets have C/Si commensurate with IDPs, and IDPs
are more carbon-rich than carbonaceous chondrites (Fig.~\ref{fig:c2s_plot}).  
Two sun-grazing comets from the Kreutz family of comets, C/2003~K7 and
C/2011~W3~(Lovejoy), have silicate-rich dust and fall in the carbonaceous chondrites (CC) range
\citep{2015PNAS..112.8965B,2013ApJ...768..161M,2010ApJ...713L..69C}. 

\citet{2017A&A...606A..16G}, \citet{2018A&A...609A..65D, 2013Icar..224..243D} and other authors
suggest that there was a carbon gradient in the early solar system. The comet C/Si values supports 
this contention of gradient in the carbon with heliocentric distance of formation. 
Commensurate with these results, CONSERT on \textit{Rosetta/Philae} suggest comets 
are a large carbon reservoir given the nucleus' permittivity and density constraints on the dust 
composition in the nucleus \cite{2016MNRAS.462S.516H}, which agrees within uncertainties with the 
average specific density of dust particles in the comet C/2013~US$_{10}$ (Catalina)'s comae. 
The existence of a carbon gradient in solar systems also is bolstered by the C/Si ratios of IDPs.  

Destruction of carbon occurred in inner disk, which is the long-standing ``carbon deficit 
problem'' \citep{2015PNAS..112.8965B, 2010ApJ...710L..21L}.  Disk modelers are 
working to predict the carbon depletion gradient with complex chemical networks 
\citep{2019ApJ...870..129W}.  Another model investigates removal of carbon through 
oxidation and photolysis when particles are transported to the exposed upper disk layers 
but radial transport erases signatures unless other mechanisms quickly destroy carbon like 
flash heating from FU Ori outbursts or mechanisms prevent replenishment of the inner 
disk such as sustained particle drift barrier, i.e., a gap opened by the formation of a giant 
planet. \citet{2018A&A...618L...1K} argue that ``a sustained drift barrier or strongly reduced 
radial grain mobility is necessary to prevent replenishment of carbon from the outer 
disk [to the inner disk]."

Heat and/or high oxygen fugacity conditions in the inner protoplanetary disk can convert carbon 
from its incorporation in refractory particles to carbon in gas phase CO or CO$_{2}$.  
As discussed (\S\ref{sec:ss-endemic-c}), particle temperatures above $\sim$823~K 
can destroy aliphatic carbon. Flash heating of Mg-Fe silicates in the presence 
of carbon is a possible formation pathway for Type I chondrules \citep{1994Natur.371..136C}. 
If cometary particles can drift interior to the water evaporation front, then cometary
materials may deliver carbon to the inner protoplanetary disk. Delivery of carbon to the gas 
phase of inner disk by comet grains requires inward delivery mechanisms during
the early pebble accretion phase of disk evolution
when the motions of aggregating materials are dominated by inward pebble drift 
\citep{2020arXiv200105007A, 2019ApJ...885..118M}. Such delivery requires
 that amorphous carbon particles already be incorporated into cometary grains in addition to
 the need that the sublimation temperature of amorphous carbon be higher than water ice 
 so that the delivery of carbon particles is interior enough for carbon to become enhanced in the gas 
 phase. High carbon abundances in the gas phase are required to explain 
the poorly graphitized carbon (PGC) halos around Fe cores in two terminal \textit{Stardust} 
particles \citep{2009M&PS...44.1611W, 2017ApJ...848..113D}.  

Earth's bulk C/Si atomic ratio is much smaller and models for its core formation and 
evolution assume a carbonaceous chondrite supply of carbon was available to form the 
Earth \citep{2015PNAS..112.8965B}.  Cometary C/Si atomic ratios are much higher than 
carbonaceous chondrites. The outer disk was richer in carbon than the inner disk.  The carbon 
gradient may be another indication of planetary gaps sculpting the compositions of small 
bodies.  \citet{2019GeCoA.261..145B} hypothesize that the isotope variances of planetary
bodies, traced through meteoritic and IDPs, can be explained if there were isotopically distinct 
nebular reservoirs of non-carbonaceous and carbonaceous that were not fully mixed in the 
primordial disk of the solar system. A planetary gap created by Jupiter's formation which inhibited 
mixing between the inner and outer disk could also explain the dichotomy in between 
non-carbonaceous and carbonaceous meteorites \citep{2019E&PSL.511...44N}.  


Cometary C/Si atomic ratios highlight the ``carbon deficit'' that occurred in the inner disk and the 
dichotomy between the inner and outer disk when juxtaposed with the C/Si atomic ratios
found for the Earth and ordinary chondrites.  Furthermore, the dust composition of many comets 
demonstrates a carbon-rich reservoir existed in the regimes of comet formation that are 
pertinent to the understanding the evolution of our protoplanetary disk and the formation of the 
planets.  


\begin{figure}[!hb]
\figurenum{9}
\centering
\includegraphics[trim=1.5cm 0cm 0cm 1.65cm,clip,width=1.10\columnwidth]{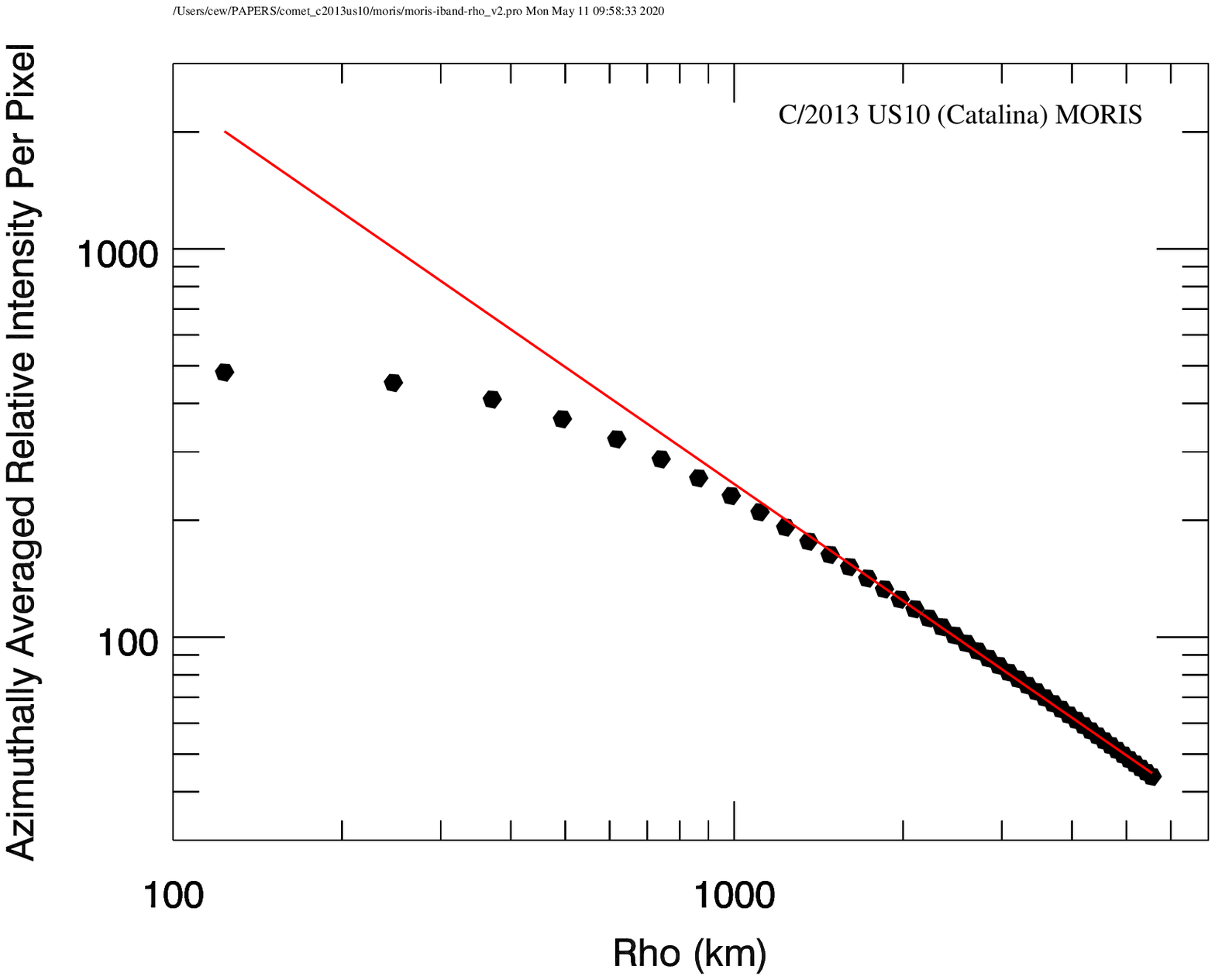}
\vspace{1.0cm}
\includegraphics[trim=1.5cm 0cm 0cm 1.65cm,clip,width=1.10\columnwidth]{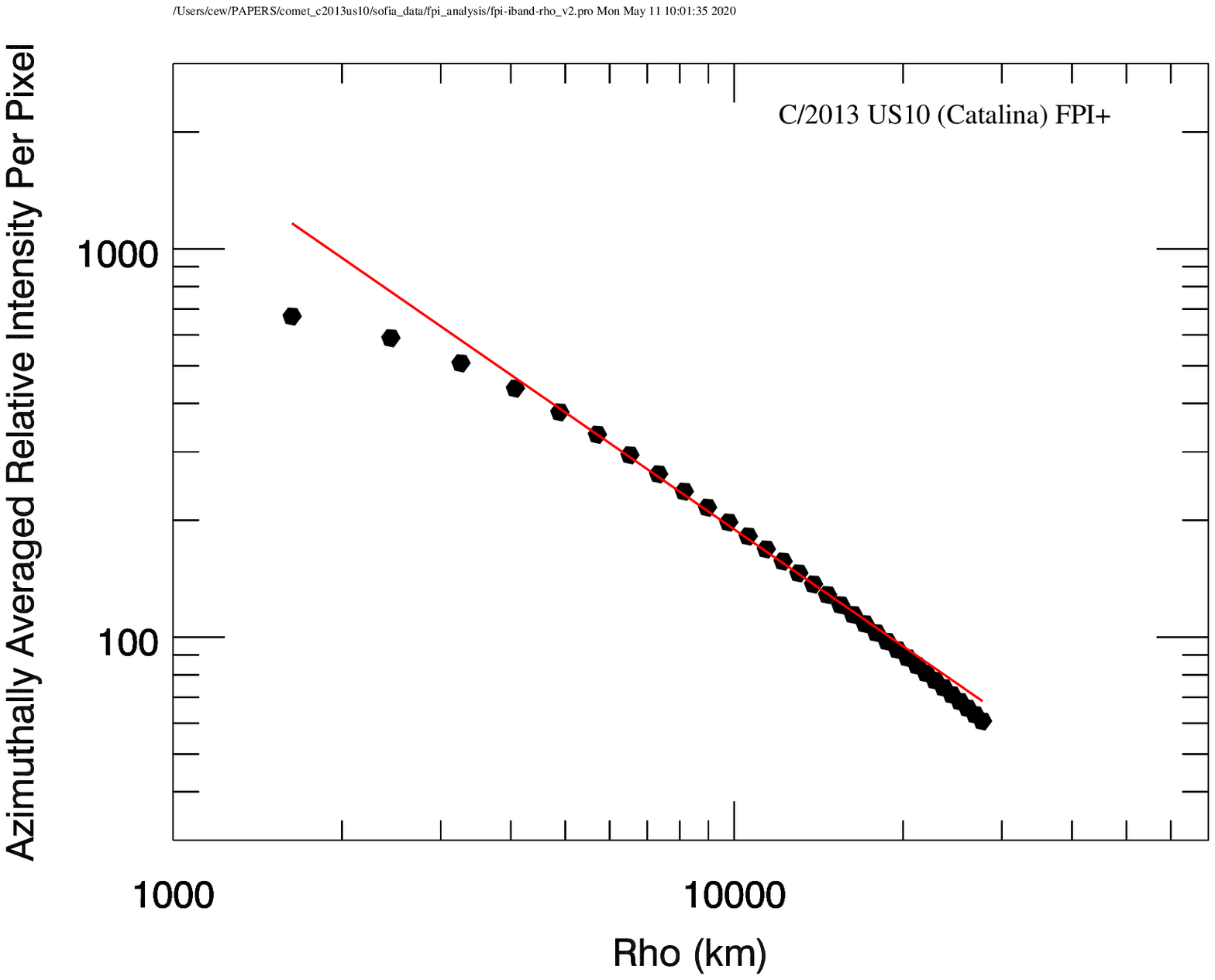}
\caption{Azimuthally averaged relative intensity per pixel
as a function of linear radius ($\rho$) in km as measured in a SDSS $i^{\prime}$-band filter
from the optical photocenter (centroid) of comet C/2013 US$_{10}$ (Catalina). 
The solid red line denotes a 1/$\rho$ profile describing a steady-state 
coma \citep[see][]{1992Icar..100..162G}. Top: the IRTF MORIS data 
obtained on 2016 Jan 11.63~UT when the phase angle was 47.80 degrees. 
Bottom: the SOFIA FPI+ data obtained on 2016 Feb 09.34~UT when the phase angle was 
33.06 degrees. Note the change in scale between the two epochs.
\label{fig:fig_i_radpro}}
\end{figure}


\subsection{Dust Production Rates}\label{sec:opt_dustproduction}

The optical spectra of comets in the $i^{\prime}$-band tends to be dominated by 
dust. However, red CN gas emission bands, CN(2,0) and CN(3,1), can present at 
redder wavelengths within the $i^{\prime}$-band 
\citep{2015SSRv..197....9C, 1991ApJ...383..356F, 1956VA......2..958S}.
Presence of these emission lines may contaminate measurements of the scattered 
light dust continuum surface brightness, and hence estimates of the dust 
production rate. Optical spectra of comet C/2013 US$_{10}$ (Catalina) obtained 
on 2015 December 18 \citep[][]{2017AJ....154..173K} show weak CN(2,0) and CN(3,1) band 
emission. However, optical spectra obtained after the epoch of the MORIS and FPI+ 
imagery in 2016 March 18 show no strong emission features redward of 7630~\AA{} 
to the $i^{\prime}$-band long wavelength cut-off \citep{2019MNRAS.484.1347H}. The 
azimuthally-averaged radial profiles of comet C/2013 US$_{10}$ (Catalina) 
derived from the MORIS and FPI+ imagery, presented in Fig.~\ref{fig:fig_i_radpro},
shows little deviation from a $1/\rho$ profile \citep{1992Icar..100..162G} at large 
cometo-centric distances consistent with a steady-state coma without significant 
CN contamination. Application of standard comet image enhancement techniques to 
these optical data \citep[see][]{2014Icar..239..168S, 2014acm..conf..462S}, reveal 
no structures in the coma such as jets or spirals at this epoch.

The dust production rate of comet C/2013 US$_{10}$ (Catalina) during the epoch of the 
BASS observations (2016 Jan 10.607 UT) was derived using the proxy quantity 
$Af\rho$ \citep{1984AJ.....89..579A}. When the cometary coma is in steady state, 
this aperture independent quantity can be parameterized as  

\begin{equation}
A(\theta)f \rho = \frac{4  \; r_{h}^{2} \;  \Delta^{2}  \; 10^{-0.4(m_{comet} - m_{\odot})}}{\rho} \, \, \rm{cm.}
\label{eqn:afrhoeqn}
\end{equation}

\noindent In this relation, $A(\theta)$ is four times the \textit{geometric} albedo 
at a phase angle $\theta$, $f$ is the filling factor of the coma, $m_{comet}$ is the 
measured cometary magnitude, $m_{\odot}$ is the apparent solar magnitude, 
derived\footnote{\url{http://classic.sdss.org/dr5/algorithms/sdssUBVRITransform.html\#vega\_sun\_colors} }
as $i^{\prime}_{\odot} = -27.002$, $\rho$ is the 
linear radius of the aperture at the comet's position (cm) and $r_{h}$(AU) and 
$\Delta$(cm) are the heliocentric and geocentric distances, respectively.

The Halley-Marcus (HM) \citep{2007ICQ....29..119M,2007ICQ....29...39M,1998Icar..132..397S}
phase angle correction\footnote{\url{http://asteroid.lowell.edu/comet/dustphase.html}}
was used to normalize $A(\theta)f \rho$ to $0\degr$ phase angle, wherein we 
adopted an interpolated value of HM = 0.3424 and 0.3946 commensurate with the 
epoch of our optical observations on 2016 Jan 11.633~UT and 2016 February 
09.340~UT, respectively. Table~\ref{tab:afrho-i-tab} reports values of
$A(0\degr)f \rho = (A({\theta})f \rho/$HM) at a selection of aperture 
sizes (distances from the comet photocenter) in the $i^{\prime}$-band. The dust 
production rate is similar to that observed in other moderately active comets, 
such as C/2012 K1 (Pan-STARRS) discussed by \citet{2015ApJ...809..181W}. 

We can roughly estimate the dust mass loss rate by taking the mass of dust observed in the 
coma inside of our aperture as the $1/\rho$ dependence of the surface brightness distribution 
indicates a steady state coma. If we adopt for the outflow velocity of 100~\micron -radii 
and larger particles which carry most of the mass a value of 
${\overline {v_{\rm dust} }} \approx 20~{\rm m~s}^{-1}$ \citep{2018MNRAS.481.1235R}, 
and assume a steady outflow of material through a spherical bubble at some distance 
$R (\rm{m})$ near the nucleus surface, the mass loss rate can be estimated as

\begin{equation} 
\dot M _{\rm dust}  \approx \frac{3\, \cdot {\rm M}_{\rm dust}  \times {\overline {v_{\rm dust} }}}{R}  
\label{eqn:dust-massloss}
\end{equation}

\noindent where $\dot M_{\rm dust}$ has units of g~s$^{-1}$. If the nucleus of 
comet C/2013 US$_{10}$ (Catalina) is comparable in size typically inferred for many comets, 1.5 km, then 
$\dot{\rm M} _{\rm dust} \approx 4 \times 10^{-3} \rm{M}_{\rm dust} \left[ {\overline {v_{\rm dust} }}/20 (\rm{m~s}^{-1})\right].$
At 1.7~au when ${\rm M} _{\rm dust}  = 4 \times10^{8}$ g (Table~\ref{tab:bf_sed_models_tab})
then $\dot M _{\rm dust} \approx 1.6 \times 10^{6}$ g~s$^{-1}$.

\citet{2012Icar..221..721F} discuss how the $A(\theta)f \rho$ can be tied to the mass production 
rate, given the HGSD parameters, computing dust mass loss rate (in kg~s$^{-1}$) assuming a 
particle density of 1~g~cm$^{-3}$ for various particle size distribution functions. Taking an average
value of $N = 3.5$, corresponding to a d$q$/d$a \sim a^{-3.5}$ which yielded 
a mass loss rate of 22.8~k~s$^{-1}$ from the detailed computations of  \citet[][see Table2]{2012Icar..221..721F} 
and using a Af$\rho$ at zero phase for C/2013 US$_{10}$ (Catalina) of $\sim 6290$~cm 
(Table~\ref{tab:afrho-i-tab}) one finds $\dot M_{\rm dust} \approx 2.4 \times 10^{6}$ g~s$^{-1}.$ 
This is comparable our latter estimate.

If we assume the density of the nucleus, which is a porous dust-ice mixture, is 
$\rho_{\rm nuc}\sim 1~\rm{g~cm}^{-3}$ \citep{2019MNRAS.482.3326F} then a 
rough estimate of the surface erosion rate from the nucleus of comet C/2013 US$_{10}$ (Catalina)
is  $\sim$1~mm~day$^{-1}$ if the entire surface is active and if the radius of the comet is $\sim$1.5~km.  
The depth of space weathering of an DN comet in the local interstellar medium might be at most 
a centimeter over the age of the solar system and this material would be shed in a
timeframe of $\ltsimeq 2$~weeks at the observed dust mass loss rate which we have 
translated to an erosion rate.  For a perspective, cumulative erosion depths for comet 
67P/Churyumov-Gerasimenko depended on the nucleus geography and solar insolation and 
from start of the \textit{Rosetta} mission until the first equinox were 6~mm to 0.1~m and to the end of the 
mission were of order 0.3~m to 4~m \citep{2020Icar..33513421C}.  

%
%


\begin{deluxetable*}{@{\extracolsep{0pt}}cccc}
\tablenum{5}
%
%
%
\tablecaption{A(0\degr)f$\rho$ Values Comet C/2013 US10 (Catalina)\label{tab:afrho-i-tab}}
\tablehead{
\colhead{Aperture}& \colhead{$\rho$} & \colhead{SDSS\, $[i]^{\prime}$} & \colhead{A(0\degr)f$\rho$}\\
\colhead{Radius} & \colhead{Radius} \\
\colhead{(arcsec)}  & \colhead{(km)}  &\colhead{(mag)}  &\colhead{(cm)}
}
\startdata
\phn1.82  & \phn990.85  & 13.040 $\pm$ 0.003  & 4618.84 $\pm$ 14.72\\
\phn3.88  & 1052.78 & 11.970 $\pm$ 0.001  & 5823.43 $\pm$ \phn0.12\\
\phn4.56  & 1238.56 & 11.771 $\pm$ 0.001  & 5950.71 $\pm$ \phn1.98\\
\phn9.12  & 2477.12 & 10.968 $\pm$ 0.002  & 6230.16 $\pm$ \phn9.93\\
13.68 & 7431.36 & 10.517 $\pm$ 0.002  & 6290.99 $\pm$ 10.96\\
18.24 & 9908.48 & 10.235 $\pm$ 0.003  & 6121.96 $\pm$ 16.48\\
\enddata
\tablecomments{\, Af$\rho$ values (computed from SDSS\, $i^{\prime}$ filter photometry) corrected 
to zero phase derived from IRTF (+MORIS) observations on 2016 Jan 11.63 UT.}
\end{deluxetable*}


The quantity $\epsilon f \rho$, \citep[see Appendix A of][]{2013Icar..225..475K},
a parameter which is the thermal emission corollary of the scattered-light 
based light $Af\rho$ was also computed using our
FORCAST broadband photometry. $\epsilon f \rho$ is defined as

\begin{equation}
\epsilon f \rho = \frac{\Delta^2}{\pi \rho} \frac{F_\nu}{B_\nu}\,\rm{(cm)},
\label{eqn:efrhoeqn}
\end{equation}

\noindent where $\epsilon$ is the effective dust emissivity,
$F_\nu$ is the flux density (Jy) of the comet within the aperture of radius
$\rho$, $B_\nu$ is the Planck function (Jy/sr) evaluated at the temperature 
$T_{\rm{c}} = T_{\rm{bb}} = 1.093 \times (278~\rm{K})\, r_{h}^{-0.5} \simeq 232.9~\rm{K},$
where $T_{c}$ is the color temperature. Derived values of $\epsilon\, f\rho$ for comet 
C/2013 US$_{10}$ (Catalina) from SOFIA photometry are presented in Table~\ref{tab:simage_phot_tab}.

\begin{figure}[!hb]
\figurenum{10}
\centering
\includegraphics[trim=1.6cm 0cm 0cm 1.85cm, clip, width=1.10\columnwidth]{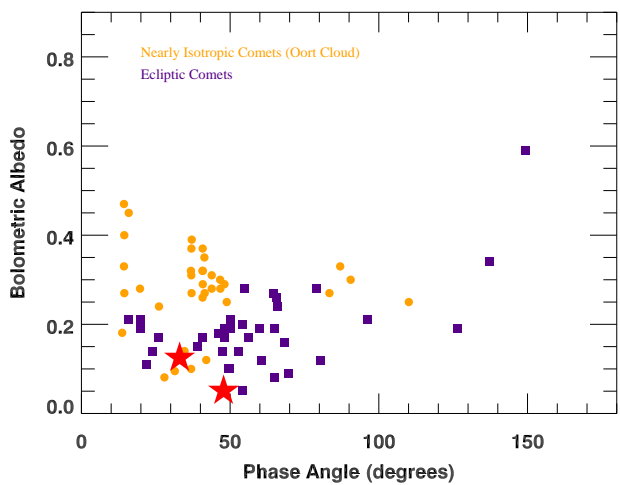} 
\caption{The bolometric albedo of ecliptic and nearly isotropic comets (Oort cloud comets) as 
a function of phase angle. Estimates for comet C/2013 US$_{10}$ (Catalina) are depicted by 
the filled red stars for both the NASA IRTF ($\simeq5$\%, 2016 Jan 10 UT) and NASA 
SOFIA ($\simeq14$\%, 2016 Feb 09 UT) observations. The bolometric albedo of C/2013 US$_{10}$ 
(Catalina)  is low compared to other isotropic (Oort cloud) comets.
\label{fig:bolo-albedo}}
\end{figure}

\subsection{Dust Bolometric Albedo}\label{sec:dust_albedo}

Our near simultaneous optical observations conducted on the same night as our measurement of 
the infrared SED of comet C/2013 US$_{10}$ (Catalina) enable us to estimate the {\it bolometric} dust 
albedo as described by \citet{2015ApJ...809..181W}. The measured albedo depends on both 
the composition and structure of the dust grains as well as the phase angle (Sun-comet observer
angle) of the observations. As the grain albedo is the ratio of the scattered light to the total
incident radiation, the thermal emission at IR wavelengths and the scattered light component 
observed at optical wavelengths are linked though this parameter.

The photometry from the $i^{\prime}$ imagery in an equivalent aperture
that corresponds to the apertures used to measure the IR SEDs 
provides an estimate of [$\lambda F_{\lambda}$]$^{max}_{scattering}$. 
An estimate of [$\lambda F_{\lambda}$]$^{max}_{IR}$ is obtained from 
a filter integrated equivalent photometric point at 10~\micron{} derived by
integrating with the observed IR SED over the band-width of the 
FORCAST F111 filter. We find that the coma of comet Oort cloud C/2013 US$_{10}$ 
(Catalina) has a low bolometric dust albedo, $A(\theta)$, of 
$\simeq5.1\pm0.1$\% at phase angle of 47.80\degr{} and 
to $\simeq13.8\pm0.5$\% at a phase angle of 33.01\degr. 

Fig.~\ref{fig:bolo-albedo} shows the derived $A(\theta)$ as a function of phase 
angle, $\theta$ for a variety of comets, where the red stars denoted the values 
for C/2013 US$_{10}$ (Catalina). At 1.3~AU, the bolometric albedo of comet C/2013 US$_{10}$ (Catalina)
is likely measuring the reflectance properties of the refractory particles because ice grains have 
very short lifetimes at this heliocentric distance \citep{2006Icar..180..473B, 2018ApJ...862L..16P}.  
Reflectance of individual refractory particles from the coma of comet 
67P/Churyumov-Gerasimenko as measured by \textit{Rosetta} COSIMA/Cosicope are 
from 3\% to 22\% at 650~nm \citep{2017MNRAS.469S.535L, 2020P&SS..18204815L}, which 
spans the range of bolometric albedos measured for comet comae.  

\section{CONCLUSION}\label{sec:the-end}

Mid-infrared $6.0 \ltsimeq \lambda(\mu\rm{m}) \ltsimeq 40$ spectrophotometric
observations of comet C/2013 US$_{10}$ (Catalina) at two temporal epochs 
yielded an inventory of the refectory materials in the comet's coma 
and their physical characteristics through thermal modeling analysis. 
The coma of C/2013 US$_{10}$ (Catalina) has a high abundance of submicron-radii  moderately
porous (fractal porosity $D = 2.727$) carbonaceous amorphous grains with a
silicate-to-carbon mass ratio $\ltsimeq 0.9$. This comet also exhibited a weak 10~\micron{} silicate 
feature. 

Comet C/2013 US$_{10}$ (Catalina) is an example of subset of comets with weak silicate 
features that are definitively shown to have low silicate-to-carbon ratios for the submicron 
grain component (as deduced from thermal model analysis of the spectral energy distributions), 
that is, they are carbon-rich. Their thermal emission is dominated by warmer particles that are 
significantly more absorbing at UV-near-IR wavelengths than silicates. The spectral grasp of 
SOFIA (+FORCAST) provided a constraint that required the presence of amorphous carbon 
as a dominate constituent of the coma particle population (submicron dust) as silicate 
particles cannot provide the lack of contrast above blackbody emission at far-infrared 
wavelengths. The surface area of the thermal emission is  dominated by the smaller grains 
and for the silicates, the smaller grains produce resonances 19.5, 23.5, 27.5~\micron{}
not evident in the spectrum of comet C/2013 US$_{10}$ (Catalina), which is a puzzle. 

A dark refractory carbonaceous material darkens and reddens the surface of the nucleus of 
67P/Churyumov-Gerasimenko. Comet C/2013 US$_{10}$ (Catalina) is carbon-rich. Analysis of 
comet C/2013 US$_{10}$ (Catalina) grain composition and observed infrared spectral features 
compared to interplanetary dust particles, chondritic materials, and \textit{Stardust} samples 
suggest that the dark carbonaceous material is well-represented by the optical properties of 
amorphous carbon. We argue that this dark material is endemic to comets.

The C/Si atomic ratio of comets in context with that derived from studies 
of interplanetary dust particles, micrometeroites, and \textit{Stardust} samples suggest that a 
carbon gradient was present in the early solar nebula. As we observe more comets, and 
especially take the opportunities to observe dynamically new comets with SOFIA, the James 
Webb Space Telescope and other capabilities, a significant subset of comets which are 
carbon-rich likely will arise providing important constraints on newly 
proposed interpretations of disk processing in the primitive solar system.

\acknowledgments
Based in part on observations made with the NASA/DLR Stratospheric Observatory for 
Infrared Astronomy (SOFIA). SOFIA is jointly operated by the Universities Space Research 
Association, Inc. (USRA), under NASA contract NNA17BF53C, and the Deutsches SOFIA Institut 
(DSI) under DLR contract 50 OK 0901 to the University of Stuttgart. Financial support for 
this work was provided by NASA through award  SOF 04-0010 and NASA PAST 
grant 80NSSC19K0868. The authors wish to thank Dr.~Aigen Lee for informative discussion regarding
carbonaceous materials and there relevance to interpreting astronomical spectra as well as
Dr. Jeff Cuzzi and the NASA Ames research group for their keen insights into disk transport 
models. The authors also express gratitude for the two anonymous referees' very careful reading of the 
manuscript and their numerous suggestions and comments that enhanced the final narrative.

\facilities{NASA SOFIA (FORCAST/FPI+), NASA IRTF (BASS/MORIS)}

\software{IRAF \citep{1986SPIE..627..733T,1993ASPC...52..173T}, IDL, 
JPL Horizons \citep{1996DPS....28.2504G}, Aperture Photometry Tool (APT) \cite{2012PASP..124..737L} }

\appendix
\section{Tables of Revised Thermal Models}

As described in the text (\S\ref{sec:ss-nurhoc}) we have adopted a value for 1.5~g~cm$^{-3}$ for the 
specific density of amorphous carbon, $\rho_{s}(\rm{ACar})$, in our thermal models. In 
early work, we employed a higher specific density of 2.5~g~cm$^{-3}$. In order to compare the 
atomic carbon-to-silicate ratios consistently thermal models for all SOFIA observed comets included 
in this analysis were modeled or remodeled with a common value of $\rho_{s}(\rm{ACar}) = 1.5$~g~cm$^{-3}$. 
Tables for 
comets C/2012 K1 (Pan-Starrs) \citep[see][]{2015ApJ...809..181W}, C/1995 O1 (Hale-Bopp)
\citep[see][]{2002ApJ...580..579H}, and C/2013 X1 (Pan-STARRS) and C/2018 W2 (Africano) 
\citep[][]{Woodward2020...inprep} are given for completeness.

\setcounter{table}{5}
%



\begin{deluxetable*}{@{\extracolsep{0pt}}lcc}
%
%
%
\tablecaption{Derived Grain Composition of Comet C/2012 K1 (Pan-STARRS)\tablenotemark{a}
\label{tab:nu_deh_sed_model_15_c12k1_tab}}
\tabletypesize{\small}
\tablehead{
&\\[-8pt]
   \multicolumn1l{\underbar{SOFIA (+FORCAST)}\tablenotemark{b}} \\
 &                             & \colhead{Relative} \\
 &                             & \colhead{Mass} \\
 &                             & \colhead{Sub-\micron} \\
\multicolumn1l{Thermal Model SED Details} & \colhead{($N_{p}\times 10^{20}$)\,\tablenotemark{c}} & \colhead{Grains}
 }
\startdata
\underbar{Dust Components}\\[2pt]
Amorphous pyroxene   & $0.658^{+ 0.084}_{- 0.133}$ & $0.398^{+ 0.077}_{- 0.085} \times 10^{7}$ \\
Amorphous olivine    & $0.000^{+ 0.085}_{- 0.000}$ & $0.000^{+  0.054}_{- 0.000}$ \\
Amorphous carbon     & $1.656^{+ 0.056}_{- 0.057}$ & $0.455^{+ 0.064}_{- 0.050}$ \\    
Crystalline olivine  & $0.181^{+ 0.161}_{- 0.165}$ & $0.147^{+ 0.104}_{- 0.133}$ \\  
Crystalline pyroxene & 0.000 & 0.000 \\[2pt]
\tableline
\underbar{Resultants}\\[2pt]
Total mass sub-\micron{} grains (gm) $\times 10^{8}$  & $3.727^{+   0.426}_{-  0.449}$ & \nodata \\
Amorphous silicate dust fraction & $0.398^{+   0.080}_{-   0.071}$ & \nodata \\    
Crystalline silicate dust fraction & $0.147^{+   0.104}_{-   0.133}$ & \nodata \\ 
Silicate to Carbon ratio\tablenotemark{$\dagger$}              & $1.198^{+   0.271}_{-   0.269}$ & \nodata \\
Crystalline silicate mass to total silicate mass\tablenotemark{d} & $0.270^{+  0.160}_{-   0.241}$ & \nodata \\
$a_{p}$(\micron)\tablenotemark{e} & 0.7  \phn \phn \phn \phn \phn \phn & \nodata \\
Fractal porosity ($D$)  & 2.857  \phn \phn \phn \phn & \nodata \\[2pt]
\tableline
\underbar{Other Parameters}\\[2pt]
Hanner Grain-Size Distribution M : N &22.2 : 3.7  \phn \phn \phn \phn \phn \phn & \nodata \\
Reduced $\chi^2_\nu$ & $1.06$ \phn \phn \phn \phn \phn \phn & \nodata \\
Degrees of freedom & 146 \phn \phn \phn \phn \phn \phn & \nodata \\ 
\enddata
\tablecomments{}
\tablenotetext{a}{Uncertainties represent the 95\% confidence level.}
\tablenotetext{b}{Comet on 2014 Jun 06 UT $r_{\rm h} = 1.70$ au, $\Delta = 1.71$ au.}
\tablenotetext{c}{Number of grains, $N_{p}$, at the peak ($a_{p}$) of the Hanner grain size distribution (GSD).}
\tablenotetext{d}{$f_{cryst} \equiv m_{cryst}/[m_{amorphous} + m_{cryst}]$ where $m_{cryst}$ is the
mass fraction of submicron crystals.}
\tablenotetext{e}{Peak grain size (radius) of the Hanner GSD.}
\tablenotetext{\dagger}{Ratio represents the bulk mass properties of the materials in the models.}
\end{deluxetable*}


%



\begin{deluxetable*}{@{\extracolsep{0pt}}lcc}
%
%
%
\tablecaption{Derived Grain Composition of Comet C/2013 X1 (Pan-STARRS)\tablenotemark{a}
\label{tab:nu_deh_sed_model_15_c13x1_tab}}
\tabletypesize{\small}
\tablehead{
&\\[-8pt]
   \multicolumn1l{\underbar{SOFIA (+FORCAST)}\tablenotemark{b}} \\
 &                             & \colhead{Relative} \\
 &                             & \colhead{Mass} \\
 &                             & \colhead{Sub-\micron} \\
\multicolumn1l{Thermal Model SED Details} & \colhead{($N_{p}\times 10^{20}$)\,\tablenotemark{c}} & \colhead{Grains}
 }
\startdata
\underbar{Dust Components}\\[2pt]
Amorphous pyroxene & $0.000^{+ 0.303}_{- 0.000}$ & $0.000^{+ 0.268}_{- 0.000}$ \\
Amorphous olivine & $0.213^{+ 0.071}_{- 0.191}$ & $0.214^{+  0.119}_{- 0.195}$ \\
Amorphous carbon & $1.197^{+ 0.064}_{- 0.066}$ & $0.545^{+ 0.168}_{- 0.149}$ \\    
Crystalline olivine & $0.138^{+ 0.212}_{- 0.138}$ & $0.241^{+ 0.213}_{- 0.241}$ \\  
Crystalline pyroxene & $0.643^{+ 0.004}_{- 0.004}$ & $0.046^{+ 0.001}_{- 0.001}$ \\[2pt]
\tableline
\underbar{Resultants}\\[2pt]
Total mass sub-\micron{} grains (gm) $\times 10^{8}$  & $1.567^{+   0.588}_{-  0.353}$ & \nodata \\
Amorphous silicate dust fraction & $0.214^{+   0.163}_{-   0.099}$ & \nodata \\    
Crystalline silicate dust fraction & $0.241^{+   0.213}_{-   0.241}$ & \nodata \\ 
Silicate to Carbon ratio\tablenotemark{$\dagger$}              & $0.834^{+   0.689}_{-   0.433}$ & \nodata \\
Crystalline silicate mass to total silicate mass\tablenotemark{d} & $0.530^{+  0.260}_{-   0.530}$ & \nodata \\
$a_{p}$(\micron)\tablenotemark{e} & 0.6  \phn \phn \phn \phn \phn \phn & \nodata \\
Fractal porosity ($D$)  & 2.727  \phn \phn \phn \phn & \nodata \\[2pt]
\tableline
\underbar{Other Parameters}\\[2pt]
Hanner Grain-Size Distribution M : N &18.5 : 3.7  \phn \phn \phn \phn \phn \phn & \nodata \\
Reduced $\chi^2_\nu$ & $0.5806$ \phn \phn \phn \phn \phn \phn & \nodata \\
Degrees of freedom & 147 \phn \phn \phn \phn \phn \phn & \nodata \\ 
\enddata
\tablecomments{}
\tablenotetext{a}{Uncertainties represent the 95\% confidence level.}
\tablenotetext{b}{Comet on 2016 Jul 13 UT $r_{\rm h} = 1.80$ au, $\Delta = 1.02$ au.}
\tablenotetext{c}{Number of grains, $N_{p}$, at the peak ($a_{p}$) of the Hanner grain size distribution (GSD).}
\tablenotetext{d}{$f_{cryst} \equiv m_{cryst}/[m_{amorphous} + m_{cryst}]$ where $m_{cryst}$ is the
mass fraction of submicron crystals.}
\tablenotetext{e}{Peak grain size (radius) of the Hanner GSD.}
\tablenotetext{\dagger}{Ratio represents the bulk mass properties of the materials in the models.}
\end{deluxetable*}


%



\begin{deluxetable*}{@{\extracolsep{0pt}}lcc}
%
%
%
\tablecaption{Derived Grain Composition of Comet C/2018 W2 (Africano)\tablenotemark{a}
\label{tab:nu_deh_sed_model_15_c18w2_tab}}
\tabletypesize{\small}
\tablehead{
&\\[-8pt]
   \multicolumn1l{\underbar{SOFIA (+FORCAST)}\tablenotemark{b}} \\
 &                             & \colhead{Relative} \\
 &                             & \colhead{Mass} \\
 &                             & \colhead{Sub-\micron} \\
\multicolumn1l{Thermal Model SED Details} & \colhead{($N_{p}\times 10^{19}$)\,\tablenotemark{c}} & \colhead{Grains}
 }
\startdata
\underbar{Dust Components}\\[2pt]
Amorphous pyroxene   & $0.166^{+ 0.692}_{- 0.166}$ & $2.698^{+ 7.153}_{- 2.534} \times 10^{7}$ \\
Amorphous olivine    & $0.488^{+ 0.294}_{- 0.488}$ & $0.077^{+  0.276}_{- 0.077}$ \\
Amorphous carbon     & $0.308^{+ 0.300}_{- 0.270}$ & $0.226^{+ 0.133}_{- 0.225}$ \\    
Crystalline olivine  & $0.082^{+ 0.340}_{- 0.082}$ & $0.647^{+ 0.166}_{- 0.146}$ \\  
Crystalline pyroxene & 0.000 & 0.000 \\[2pt]
\tableline
\underbar{Resultants}\\[2pt]
Total mass sub-\micron{} grains (gm) $\times 10^{18}$  & $0.000^{+   3.719}_{-  0.000}$ & \nodata \\
Amorphous silicate dust fraction & $0.050^{+   0.166}_{-   0.050}$ & \nodata \\    
Crystalline silicate dust fraction & $0.000^{+   0.187}_{-   0.000}$ & \nodata \\ 
Silicate to Carbon ratio\tablenotemark{$\dagger$}              & $0.303^{+   0.122}_{-   0.161}$ & \nodata \\
Crystalline silicate mass to total silicate mass\tablenotemark{d} & $0.050^{+  0.224}_{-   0.050}$ & \nodata \\
$a_{p}$(\micron)\tablenotemark{e} & 0.4  \phn \phn \phn \phn \phn \phn & \nodata \\
Fractal porosity ($D$)  & 2.857  \phn \phn \phn \phn & \nodata \\[2pt]
\tableline
\underbar{Other Parameters}\\[2pt]
Hanner Grain-Size Distribution M : N &10.2 : 3.4  \phn \phn \phn \phn \phn \phn & \nodata \\
Reduced $\chi^2_\nu$ & $0.84$ \phn \phn \phn \phn \phn \phn & \nodata \\
Degrees of freedom & 134 \phn \phn \phn \phn \phn \phn & \nodata \\ 
\enddata
\tablecomments{}
\tablenotetext{a}{Uncertainties represent the 95\% confidence level.}
\tablenotetext{b}{Comet on 2019 Oct 18 UT $r_{\rm h} = 1.60$ au, $\Delta = 0.84$ au.}
\tablenotetext{c}{Number of grains, $N_{p}$, at the peak ($a_{p}$) of the Hanner grain size distribution (GSD).}
\tablenotetext{d}{$f_{cryst} \equiv m_{cryst}/[m_{amorphous} + m_{cryst}]$ where $m_{cryst}$ is the
mass fraction of submicron crystals.}
\tablenotetext{e}{Peak grain size (radius) of the Hanner GSD.}
\tablenotetext{\dagger}{Ratio represents the bulk mass properties of the materials in the models.}
\end{deluxetable*}


%



\begin{deluxetable*}{@{\extracolsep{0pt}}lcc}
%
%
%
\tablecaption{Derived Grain Composition of Comet C/1995 O1 (Hale-Bopp)\tablenotemark{a}
\label{tab:nu_deh_sed_model_15_hp_tab}}
\tabletypesize{\small}
\tablehead{
&\\[-8pt]
   \multicolumn1l{\underbar{HIFOGS + ISO Spectra}\tablenotemark{b}} \\
 &                             & \colhead{Relative} \\
 &                             & \colhead{Mass} \\
 &                             & \colhead{Sub-\micron} \\
\multicolumn1l{Thermal Model SED Details} & \colhead{($N_{p}\times 10^{22}$)\,\tablenotemark{c}} & \colhead{Grains}
 }
\startdata
\underbar{Dust Components}\\[2pt]
Amorphous pyroxene & $5.552^{+ 0.009}_{- 0.009}$ & $0.304^{+ 0.001}_{- 0.001}$ \\
Amorphous olivine & $6.334^{+ 0.005}_{- 0.005}$ & $0.347^{+  0.001}_{- 0.001}$ \\
Amorphous carbon & $3.169^{+ 0.002}_{- 0.002}$ & $0.079^{+ 0.001}_{- 0.001}$ \\    
Crystalline olivine & $3.182^{+ 0.002}_{- 0.002}$ & $0.224^{+ 0.001}_{- 0.001}$ \\  
Crystalline pyroxene & $0.643^{+ 0.004}_{- 0.004}$ & $0.046^{+ 0.001}_{- 0.001}$ \\[2pt]
\tableline
\underbar{Resultants}\\[2pt]
Total mass sub-\micron{} grains (gm) $\times 10^{8}$  & $420.133^{+   0.009}_{-  0.006}$ & \nodata \\
Amorphous silicate dust fraction & $0.651^{+   0.001}_{-   0.001}$ & \nodata \\    
Crystalline silicate dust fraction & $0.270^{+   0.001}_{-   0.001}$ & \nodata \\ 
Silicate to Carbon ratio\tablenotemark{$\dagger$}               & $11.678^{+   0.008}_{-   0.008}$ & \nodata \\
Crystalline silicate mass to total silicate mass\tablenotemark{d} & $ 0.293^{+  0.001}_{-   0.001}$ & \nodata \\
$a_{p}$(\micron)\tablenotemark{e} & 0.2  \phn \phn \phn \phn \phn \phn & \nodata \\
Fractal porosity ($D$)  & 2.857  \phn \phn \phn \phn & \nodata \\[2pt]
\tableline
\underbar{Other Parameters}\\[2pt]
Hanner Grain-Size Distribution M : N &3.4 : 3.4  \phn \phn \phn \phn \phn \phn & \nodata \\
Reduced $\chi^2_\nu$ & $11015.16$ \phn \phn \phn \phn \phn \phn & \nodata \\
Degrees of freedom & 513 \phn \phn \phn \phn \phn \phn & \nodata \\ 
\enddata
\tablecomments{}
\tablenotetext{a}{Uncertainties represent the 95\% confidence level.}
\tablenotetext{b}{Comet on 1995 Oct 11 UT $r_{\rm h} = 2.80$ au, $\Delta = 3.03$ au.}
\tablenotetext{c}{Number of grains, $N_{p}$, at the peak ($a_{p}$) of the Hanner grain size distribution (GSD).}
\tablenotetext{d}{$f_{cryst} \equiv m_{cryst}/[m_{amorphous} + m_{cryst}]$ where $m_{cryst}$ is the
mass fraction of submicron crystals.}
\tablenotetext{e}{Peak grain size (radius) of the Hanner GSD.}
\tablenotetext{\dagger}{Ratio represents the bulk mass properties of the materials in the models.}
\end{deluxetable*}


\clearpage
\bibliography{msus10-accept}{}
\bibliographystyle{aasjournal}

\end{document}